\documentclass[12pt,iopams]{iopart}
\usepackage{graphicx,color}
\usepackage{graphics,lscape,epsfig}
\usepackage{dcolumn}
\usepackage{bm}
\usepackage{epsfig,color}
\usepackage{natbib}
\usepackage{amssymb}
\newcommand{\pdag}{{\phantom{\dagger}}}
\newcommand{\be}{\begin{equation}}
\newcommand{\ee}{\end{equation}}
\newcommand{\ba}{\begin{eqnarray*}}
\newcommand{\ea}{\end{eqnarray*}}
\newcommand{\bea}{\begin{eqnarray}}
\newcommand{\eea}{\end{eqnarray}}
\newcommand{\bk}{{\mathbf{k}}}

\newcommand{\Ima}{{\Im m}}
\newcommand{\Rea}{{\Re e}}
\newcommand{\newblock}{\hskip .11em plus .33em minus .07em}

\begin{document}

\topical{Strong correlations in a nutshell}
\author{Michel Ferrero$^1$, Lorenzo De Leo$^{1,2}$, Philippe Lecheminant$^3$ and Michele Fabrizio$^{4,5}$}
\address{$^1$ Centre de Physique Th\'eorique, Ecole Polytechnique, 91128 Palaiseau Cedex, France}
\address{$^2$ Center for Material Theory, Serin Physics Laboratory, Rutgers University, 136
Frelinghuysen Road, Piscataway, NJ 08854-8019, USA}
\address{$^3$ Laboratoire de Physique Th\'eorique et Mod\'elisation, Universit\'e de Cergy-Pontoise, CNRS UMR 8089, 
2 Avenue Adolphe Chauvin, 95302, Cergy-Pontoise, France}
\address{$^4$ International
School for Advanced Studies (SISSA), and INFM-Democritos, National Simulation Center, I-34014 Trieste, Italy}
\address{$^5$ The Abdus Salam International Center for Theoretical Physics
(ICTP), P.O.Box 586, I-34014 Trieste, Italy}

\date{\today}
\begin{abstract}
We present the phase diagram of clusters made of two, three and four coupled Anderson impurities.
All three clusters share qualitatively similar phase diagrams
that include Kondo screened and unscreened regimes separated by almost critical crossover regions reflecting
the proximity to barely avoided critical points. This suggests the emergence of universal paradigms 
that apply to clusters of arbitrary size. We discuss how these crossover regions of the impurity models
might affect the approach to the Mott transition within a cluster extension of dynamical mean-field theory.
\end{abstract}
\maketitle

\tableofcontents

\begin{table}[htb]
\caption{Acronyms and main notations used in the text. }
\begin{indented}
\item[] \begin{tabular}{@{}ll}
\br
$W$ & Non-interacting electron bandwidth \\
$U$ & Hubbard on-site repulsion \\
$T_F^*$ & Quasi-particle effective Fermi temperature\\
$\Gamma$ & Impurity hybridization width \\
$T_K$ & Kondo temperature \\
$\rho(\epsilon)$ & Impurity density of states\\
$\Sigma(i\omega)$ & Impurity self-energy in Matsubara frequencies\\
MIT & Mott metal-to-insulator transition\\
DMFT &  Dynamical mean-field theory\\
NRG & Wilson's numerical renormalization group \\
CFT & Conformal field theory\\
DOS & Single-particle density of states \\
\br
\end{tabular}
\end{indented}
\end{table}

\section{Introduction}
\label{Introduction}

The Mott metal-to-insulator transition~\cite{Mott1949,Mott} emerges out of the competition
between the opposite tendencies of the electrons to delocalize throughout the lattice
in order to maximize the band-energy gain and their mutual Coulomb
repulsion which, on the contrary, tends
to suppress valence fluctuations by localizing the carriers. If
the band-energy gain, which can be identified with the ``bare''
bandwidth $W$, is small enough with respect to the short-range
Coulomb repulsion, commonly parametrized by an on-site Hubbard
$U$, and the average electron-density per site is integer, charge gets 
localized and the system is a Mott insulator; otherwise it remains metallic.

Despite its intuitive nature, the Mott phenomenon is extremely difficult to study
because it is inherently non-perturbative and because it escapes any simple
single-particle description. Those can only deal with band-insulators, characterized by an energy gap separating totally filled
from unfilled bands. The simplest example of a Mott insulator is provided by the single-band Hubbard model at half-filling,
which always has an insulating phase at sufficiently large repulsion.
Yet, in order to make this phase appear for instance in Hartree-Fock
theory, one is obliged to assume an antiferromagnetic order parameter that
doubles the unit cell so as to fulfill the necessary requirement for
a band-insulator -- an even number of electrons per unit cell.
With this trick, the Mott transition is effectively turned into a metal to band-insulator 
transition driven by magnetism. 
In reality, local moments form and eventually order as a consequence of charge localization by the Mott phenomenon.
This distinction might look pedantic since the ground state is anyway both insulating and magnetic,
but in fact it is not, as we are going to argue by the qualitative behavior of the
entropy in the Mott insulator and in the contiguous metal.

\begin{figure}
\centerline{\includegraphics[width=12cm]{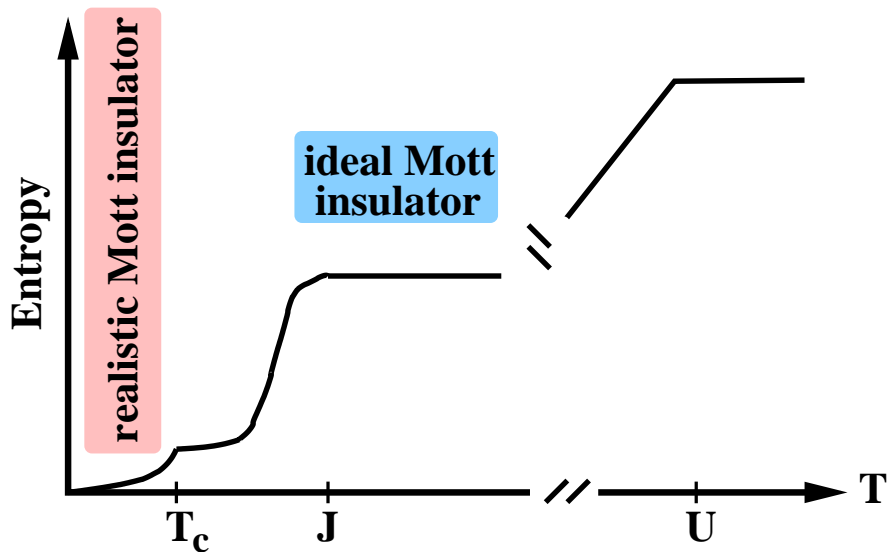}}
\caption{\label{entropyMott} Qualitative behavior of the entropy in a Mott insulator}
\end{figure}

In Fig.~\ref{entropyMott}, we sketch the typical temperature dependence of the entropy 
deep inside a Mott insulating phase, $U\gg W$.
At high temperature, $T\lesssim U$, valence fluctuations are suppressed and the local charge gets locked to some 
fixed value $n_0$. Yet, all local electronic configurations with $n_0$ electrons are thermally occupied with equal probability, 
leading to a constant entropy regime that could be identified as the \underline{ideal} Mott insulator, where    
charge degrees of freedom are frozen while all other degrees of freedom, in particular 
spin, are completely free.  However, at
some lower temperature, other energy scales come into play whose
role is to lock these additional degrees of freedom, namely to favor one or several among 
all the local electronic configurations with $n_0$ electrons. These energy scales
may include for instance the on-site Coulomb exchange, responsible for the Hund's
rules, the inter-site direct- or super-exchange, the crystal field, the coupling
to the lattice, etc. We will collectively denote these energy scales by $J$,
which may be identified as the temperature below which the entropy of the residual degrees of
freedom of the ideal Mott insulator starts to be suppressed.
Consequently, at low temperature, a \underline{realistic}
insulating phase is established, which is commonly
accompanied by a symmetry breaking phase transition at $T=T_c\leq J$, for instance a magnetic
ordering, a collective Jahn-Teller distortion, etc.  
Below $T_c$, the entropy decreases to zero as $T\to 0$, generically faster than linearly.
For instance, in the half-filled single-band Hubbard model, the \underline{ideal} Mott insulator corresponds  
to a regime in which each site is singly occupied but its spin can be with equal probability either up or down, leading to an entropy 
$\ln 2$ per site. However, below a temperature of the order of the inter-site spin-exchange, the $\ln 2$ entropy decreases  
until the system crosses a magnetic phase transition, below which its entropy vanishes according to the dimensionality of the 
system and to the dispersion relation of the spin waves. 

Recently, an amount of research activity has focused on the
possibility that different symmetry broken phases may compete in
the insulator, leading to exotic low temperature phenomena~\cite{misguich-2005-}.
Here, we will completely discard this event and concentrate on a different competition which
emerges in the metallic phase adjacent the Mott insulator.

\begin{figure}
\centerline{\includegraphics[width=12cm]{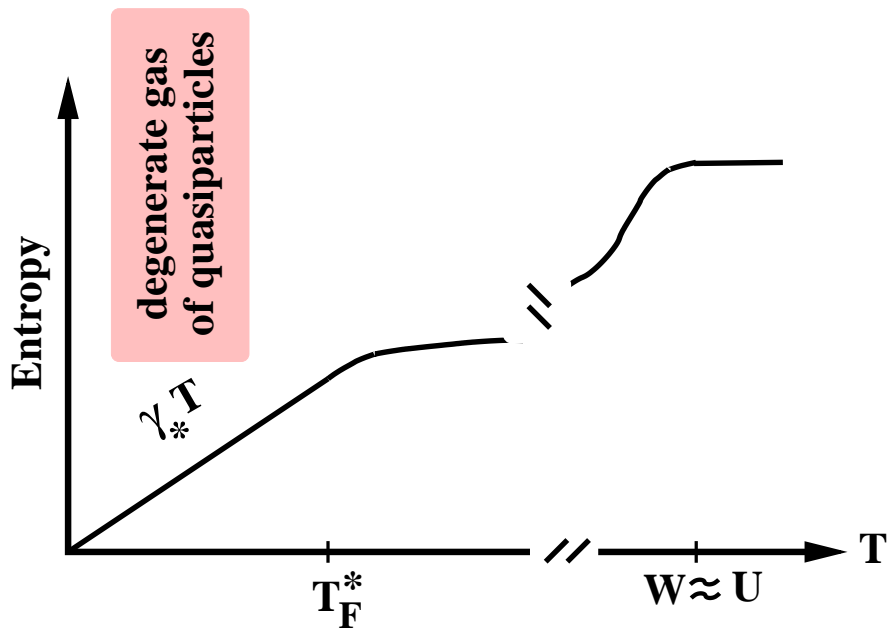}}
\caption{Qualitative behavior of the entropy of a strongly-correlated metal.}
\label{entropymetal}
\end{figure}

In Fig.~\ref{entropymetal}, we draw how the entropy versus temperature might look like 
for a strongly-correlated Fermi-liquid-like metal, assuming
that no symmetry breaking intervenes down to zero temperature.
As before, we expect that the charge entropy is, this time only partially,
reduced at some high temperature of order $U\simeq W$. The rest of it,
as well as the entropy of the other degrees of freedom, are instead suppressed by the
formation of the degenerate quasi-particle gas. This occurs below a temperature $T_F^*$, that can
be identified as the effective quasi-particle Fermi temperature. Since quasi-particles carry
the same quantum numbers as the electrons, the entropy quenching involves all degrees of freedom
at once, including the charge. Below $T_F^*$ the entropy vanishes linearly,
$S(T) \simeq \gamma_*\, T$, with a specific heat coefficient $\gamma_*$ usually larger than its
non-interacting value $\gamma \sim 1/W$.

Let us suppose that the Mott transition (MIT) were continuous and try to guess how that might happen from
the point of view of the entropy.
Obviously, since quasi-particles disappear in the Mott insulator, $T_F^*$ has to vanish at the MIT. Therefore,
sufficiently close to the MIT, the quasi-particle Fermi temperature $T_F^*$ must become smaller than $J$.
When this happens we should expect, by continuity with the insulating side, that part of the spin entropy
gets suppressed already at temperatures of order $J$, above the onset of Fermi-degeneracy. This amounts to 
some kind of pseudo-gap opening above $T_F^*$, which is at odds with the conventional Landau-Fermi-liquid theory. 
One way out, apart from a first-order MIT, is that something new occurs
when $T_F^* \simeq J$. Indeed, the presence of $J$ provides the metallic phase
with an alternative mechanism to freeze spin
degrees of freedom independently of the charge ones, a mechanism that becomes competitive
with the onset of a degenerate quasi-particle gas when $T_F^* \simeq J$. This competition is likely
to lead to an instability of the Landau-Fermi-liquid towards a low-temperature symmetry-broken phase 
(this happens in the insulating side), but may also signal a real break-down of Fermi-liquid theory.

Notice that, unlike the competition between different symmetry broken Mott-insulating
phases, which requires a fine tuning of the Hamiltonian parameters
that may only accidentally occur in real materials, this other type 
of competition -- whose effects have not been discussed in the literature before to the extent we believe they deserve -- 
should be encountered whenever it is possible to
move gradually from a Mott insulator into a metallic phase, for
instance by doping or by pressure. We also know several examples
where this competition is argued to be at the origin of interesting
phenomena. For instance, in heavy fermion materials the Kondo
effect, favoring the formation of a coherent band of heavy
quasi-particles, competes with the RKKY interaction (for a comprehensive review 
see e.g. Ref.~\cite{Hewson}). Here this
competition is supposedly the key to understand the anomalies
which appear at the transition between the heavy fermion
paramagnet and the magnetically ordered phase~\cite{millis-1993,coleman-2001-13,si-2004-}.

\subsection{Competing screening mechanisms in Anderson impurity models}
\label{Intro:A}

The heavy-fermion example is a particularly pertinent one to 
introduce the subject of this Topical Review. Indeed, the competition
between Kondo effect and RKKY coupling has interesting
consequences not only in the periodic Anderson model but already
at the level of Anderson impurity models. 

For instance, the phase
diagram of two spin-1/2 impurities coupled to a conduction bath
and mutually by a direct antiferromagnetic exchange has two limiting
regimes: one where each impurity is independently Kondo screened
by the conduction electrons; and another where the 
exchange locks the impurity spins into a singlet
state, now transparent to the conduction electrons. Under
particular circumstances -- the two involved scattering channels must be independently coupled each to one impurity --
these two regimes are separated by a
quantum critical point, at which non-Fermi liquid behavior
emerges~\cite{Jones87,Jones88,Jones89,Affleck92PRL,Affleck95}.

The phase diagram grows richer when one consider three
antiferromagnetically coupled spin-1/2
impurities~\cite{paul-1996-,ingersent-2005-95}. Here, besides a
Kondo screened regime, there are other phases where the direct
exchange prevails, but is unable to fully quench all impurity
degrees of freedom. This leads to stable non-Fermi liquid phases
analogous to overscreened multi-channel Kondo
models~\cite{Affleck:1990by,Affleck:1990iv}. These impurity-cluster models are interesting not only as simple attempts
towards an understanding of the fully periodic Anderson model, but also
because compact cluster of impurities are
achievable experimentally by adsorbing atoms on metallic surfaces.
Trimers of Cr atoms have already been realized on gold
surfaces~\cite{Jamneala2001mc}, which has actually motivated the
most recent theoretical activity on impurity
trimers~\cite{Kudasov2002u,savkin-2005-94,lazarovits-2005-95,ingersent-2005-95}.
In this context, the major task is to identify those phases which
are stable towards perturbations generically present on
metallic surfaces. Therefore the quantum critical points
that separate stable phases are of minor interest from an experimental point of view, as they require such 
a fine tuning that is extremely unlikely to occur in physical systems.

\subsection{Impurity models and Dynamical Mean Field Theory}

Unstable critical points arise when the competition between Kondo screening and
RKKY coupling is maximum. This is nothing but the impurity counterpart of the situation
in which $T_F^* \simeq J$ we previously met in connection with the Mott transition.
This weak analogy turns into an actual equivalence within the so-called
dynamical mean-field theory (DMFT)~\cite{George96}, 
the quantum analogue of classical mean-field theory which, like the latter,
is exact for infinite coordination lattices. In this limit, the
single-particle self-energy becomes fully local but maintains a non-trivial
time-dependence, obtained within DMFT by solving an auxiliary single-impurity Anderson model
that is designed so as to have an impurity self-energy that coincides with the local self-energy of
the lattice model. This requirement translates into an impurity
model identified by the same local interaction as the
lattice model and by a coupling to a conduction bath that must 
be self-consistently determined. The single-site formulation of DMFT has provided
a lot of useful information about the Mott transition {\sl per se}, disentangled from magnetism
or whatever symmetry breaking occurs in the insulating state. However, 
even though single-site DMFT can account in a Hartree-Fock manner for simple magnetic phases on bipartite lattice,  
it is inadequate to study our anticipated competition. For instance, it misses precursor effects in the 
paramagnetic phase close to the magnetic phase transition, caused by inter-site
processes which disappear in infinite coordination lattices.
For this reason, several extensions of DMFT have been recently proposed to include
short-range spatial correlations~\cite{Kotliar2001spb,Maier2005,senechal-2000-84,potthoff-2003-91,Lichtenstein2000k}.
In these novel versions, the lattice model is mapped onto \underline{a cluster} of
Anderson impurities, subject to self-consistency conditions. 

The physics of the Anderson impurity model turned out to be a precious guideline
to interpret single-site DMFT results~\cite{George96}. Similarly, we expect
that the preliminary knowledge of the general properties of impurity clusters is useful, perhaps even 
necessary, in connection with any cluster-extension of DMFT. However, apart from few
exceptions~\cite{Jones87,paul-1996-}, little is known about impurity clusters.
In addition, since models of impurity clusters involve many intra- and inter-impurity energy scales, it is not
{\sl a priori} evident that there should be  a common interpreting framework
like the Kondo physics in the single-impurity case.

This is actually the purpose of this Topical Review. Specifically, we are going to present the
phase diagram of the simplest among impurity clusters, namely dimers, trimers and tetramers, that could be used to
implement a cluster DMFT calculation on strongly correlated models. 
Besides our main objective to identify the common features among different clusters,
which should presumably play the most significant role in a DMFT approach, we will also try to argue how much of the
impurity cluster physics might survive the DMFT self-consistency, hence the possible consequences on the phase diagram
of the lattice models.
Needless to say, the interest in impurity clusters
goes beyond its possible relevance to strongly correlated models near a Mott transition.
As we previously mentioned, these clusters may be experimentally realized
on metallic surfaces or, eventually, by arranging quantum dots in
proper geometries. Moreover,
these models represent a theoretical challenge by themselves, which
requires the full machinery of Wilson's numerical renormalization
group (NRG)~\cite{Wilson75,Krishnamurthy80i,Krishnamurthy80ii} and conformal
field theory (CFT)~\cite{DiFrancesco} for a detailed comprehension.

\bigskip

Before entering into the details of our calculations, it is worth briefly
presenting the physical idea that guided this work.
First of all, let us recall some basic facts of the
single-site DMFT mapping onto impurity models. Within this mapping, the quasi-particle
effective Fermi temperature, $T_F^*$, translates into the Kondo temperature, $T_K$, of the impurity model.
The self-consistency condition causes $T_K$ to vanish at a finite
value of $U$, which signals, in the lattice counterpart, the
onset of the Mott transition. This also implies that the metallic
phase just prior to the Mott transition translates into an
Anderson impurity model deep inside the Kondo regime, with a very
narrow Kondo resonance and pre-formed Hubbard side-bands~\cite{George96}.
The same behavior should occur even when dealing with a cluster of impurities,
which should therefore translate into a cluster of Kondo impurities.
The novelty stems from the other energy scales which we collectively denoted as
$J$, and that take care of quenching in the Mott insulator the degrees of
freedom other than the charge. Indeed, near the Mott transition, $J$ translates into
additional processes, like for instance a direct exchange between the impurity-spins, which tend to remove,
completely or partially, the degeneracy of the cluster. Consequently, $J$
competes with the Kondo effect, an agent that takes more advantage the
more degenerate the impurity-cluster ground state.
\begin{figure}
\centerline{\includegraphics[width=12cm]{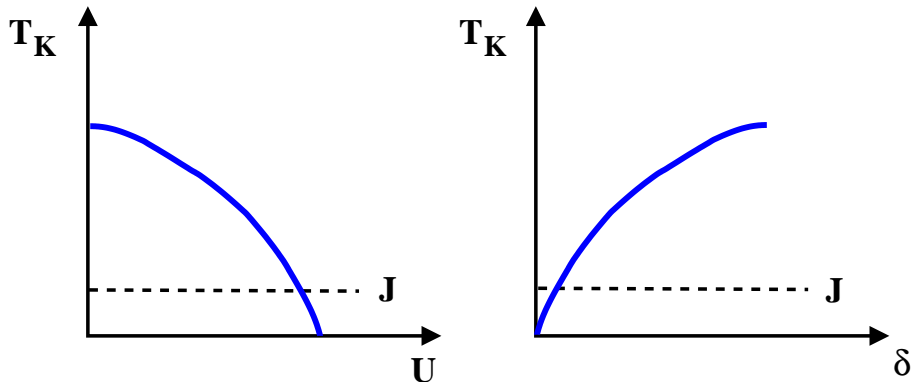}}
\caption{\label{Fig1} Behavior versus $U$ and doping $\delta$ 
of the quasiparticle Fermi temperature, $T_F^*$, which translates within DMFT into the
Kondo temperature $T_K$ of the effective impurity model. $J$ is an effective intra-cluster energy scale.}
\end{figure}

We notice that this competition is always active in impurity clusters, 
while it is commonly absent in single-impurity models except in 
multi-orbital cases~\cite{Capone02,Capone04} whose physics is in fact close to clusters. We believe that this
additional ingredient is precisely the common denominator of all impurity-cluster models, which endows them
with the capability of providing a more faithful description of a realistic
Mott transition within DMFT.

Indeed, in the presence of the intra-cluster coupling $J$, the approach to the Mott transition
changes as qualitatively shown in Fig.~\ref{Fig1}, with a Kondo
temperature smoothly decreasing from its initial value $W$ as
$U/W$ increases and becoming of order $J$ just before the transition. 
Analogously, see also Fig.~\ref{Fig1}, starting 
from the Mott insulator and doping it, $T_K$ will smoothly increase
from its value $T_K=0$ at zero doping, until it will again cross a
value of order $J$. In other words, any impurity cluster
should experience, within DMFT, two different regimes. The first, when $T_K\gg J$, in which full Kondo
screening takes place and the impurity density of states
displays the usual Kondo-resonance. In the lattice model, this regime translates into a conventional correlated
metal. The second, when $T_K\ll J$, particularly close to the Mott transition, in which
no or only partial Kondo screening occurs. Here the impurity density of states is pseudo-gaped at the
chemical potential. As we will show, these two regimes of the impurity cluster are generically separated by
an almost critical crossover-region that reflects the proximity to a true quantum critical point.
How do the unscreened phase and the almost critical crossover-region of the impurity cluster translate
in the lattice model? The answer to this question would be simple if a true impurity critical point
existed, as discussed in Refs.~\cite{DeLeo03f} and \cite{DeLeo04f}. Indeed, near a
critical point, the impurity model displays strongly enhanced
local susceptibilities, equivalently enhanced local irreducible four-leg vertices,
in several instability channels. Within DMFT, the irreducible four-leg
vertices, which enter the Bethe-Salpeter equations, coincide with the
local ones~\cite{George96}. Therefore, it is reasonable to argue that,
after full DMFT self-consistency is carried out, these local
instabilities may turn into symmetry-breaking bulk-instabilities that correspond to the same instability channels
of the impurity critical point. However, establishing which one of these symmetry breakings really occurs
requires full DMFT calculations, as it depends on other details, like for instance the nesting of the
Fermi surface. These speculations have been tested with success by a DMFT analysis of
a two-orbital Hubbard model~\cite{Capone04}. Although these criticalities
are, rigorously speaking, avoided in impurity-cluster models pertinent to DMFT, still we believe that these
models approach a critical point so closely that the physics does not change qualitatively.

In the following, we will describe in succession the case of dimer, trimer and tetramer clusters, and, in spite of their 
obvious differences, we will identify the universal aspects of the competition discussed above.

\section{The impurity dimer}
\label{The impurity dimer}
The simplest impurity cluster that is relevant to DMFT is a dimer described by the Hamiltonian
\bea
\mathcal{H} &=& \sum_{a=1}^2\,\sum_{\bk\sigma} \, \epsilon_\bk \, c^\dagger_{a\,\bk\sigma} \, c^\pdag_{a\,\bk\sigma} \,
  + \, \sum_{\bk\sigma} \, t_{\perp\,\bk}\,\left( c^\dagger_{1\,\bk\sigma} \, c^\pdag_{2\,\bk\sigma} + H.c. \right) \nonumber \\
  && - \sum_{a=1}^2\,\sum_{\bk\sigma} \,\left(V_\bk\,c^\dagger_{a\,\bk\sigma} d^\pdag_{a\,\sigma} + H.c.\right)\label{AIM-dimer}\\
  && -t_\perp\,\sum_\sigma \Big(d^\dagger_{1\,\sigma} d^\pdag_{2\,\sigma}
  + H.c.\Big)+ \frac{U}{2}\sum_{a=1}^2 \, (n_a - 1)^2,\nonumber
\eea
where $c^\dagger_{a\,\bk\sigma}$ creates a conduction electron in channel $a=1,2$ with
momentum $\bk$, energy $\epsilon_\bk$, measured with respect to the chemical potential, and spin $\sigma$,
while $d^\dagger_{a\,\sigma}$ is the creation
operator at the impurity site $a=1,2$ with spin $\sigma$, $n_a = \sum_\sigma\,
d^\dagger_{a\,\sigma}  d^\pdag_{a\,\sigma}$ being the occupation number.
This model describes two Anderson impurities, each hybridized with its own conduction bath
and in turn coupled to the other impurity by a single-particle hopping $t_\perp$.
Since within cluster DMFT the self-consistent baths 
must mimic the effects of the rest of the lattice on the two sites of the dimer,
also the two baths are coupled by a hybridization $t_{\perp\,\bk}$.
The role of the inter-bath hybridization is to generate a frequency-dependent contribution to the
inter-impurity hopping that, together with $t_\perp$, produce off-diagonal elements $a \not = b$ to
the impurity Green's function
\[
\mathcal{G}_{ab}(\tau) = -\langle T_\tau\Big(d^\pdag_{a\,\sigma}(\tau)\,d^\dagger_{b\,\sigma}\Big)\rangle.
\]
In fact, any coupling among the baths transfers into a coupling among the impurities and vice-versa, 
apart from the frequency dependence that can be anyway neglected in the asymptotic low-frequency regime.
For this reason, in what follows we will indifferently refer to inter-bath or to inter-site depending upon the context.  

\begin{figure}
\centerline{\includegraphics[width=12cm]{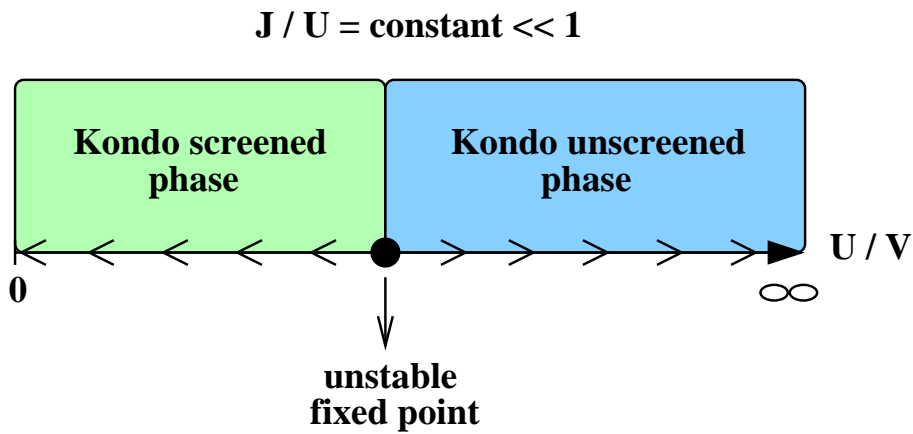}}
\caption{\label{dimer-phd} Phase diagram of the dimer model (\ref{Ham-dimer}) as function of $U/\Gamma$ at
fixed $J/U\ll 1$. }
\end{figure}

Close to the Mott transition
the effective impurity model resides well inside the Kondo regime, where $U\gg V_\bk,t_\perp$.
Here, the model can be mapped via a Schrieffer-Wolff transformation onto two spin-1/2 impurities that, up to order $1/U$,
are coupled to the two conduction baths by a Kondo exchange
\be
J_K = \frac{8}{U}\,\sum_\bk\, \left|V_\bk\right|^2 \, \delta\left(\epsilon_\bk\right)
\label{JK}
\ee
and together by an antiferromagnetic
$J = 4t_\perp^2/U$. This means that the spectral distribution of the
inter-impurity hybridization
\[
\sum_\sigma\, \langle \Big(d^\dagger_{1\,\sigma} d^\pdag_{2\,\sigma}
+ H.c.\Big)\rangle = -4\,\int_{-\infty}^0 \frac{d\omega}{\pi}\, \Ima\, \mathcal{G}_{12}(\omega) ,
\]
is transferred to high energy, and what remains at low energy is mainly the
exchange $J$. Within the DMFT self-consistency scheme, we should then expect that also the
direct hybridization among the baths, $t_{\perp\,\bk}$, behaves
similarly, which suggests that one could start the analysis with the large $U$-limit of the Hamiltonian
\bea
\mathcal{H} &=& \sum_{a=1}^2\Bigg[\sum_{\bk\sigma}
\,\epsilon_\bk\, c^\dagger_{a\,\bk\sigma} c^\pdag_{a\,\bk\sigma} -
\left(V_\bk\,c^\dagger_{a\,\bk\sigma} d^\pdag_{a\,\sigma} + H.c.\right)\Bigg]\nonumber\\
&+& \frac{U}{2}\sum_{a=1}^2 \, (n_a - 1)^2 + J\, \mathbf{S}_1\cdot\mathbf{S}_2 \nonumber \\
&& \equiv \sum_{a=1}^2\,\mathcal{H}_{a}^K + J\, \mathbf{S}_1\cdot\mathbf{S}_2,
\label{Ham-dimer}
\eea
where $\mathbf{S}_a$ is the spin-density operator of impurity $a=1,2$, plus a weak inter-bath hybridization
to be considered as a perturbation. This does not at all imply that the latter is irrelevant.
Rather, as we are going to discuss, this hybridization turns out to be a relevant perturbation. It only means that
this perturbation becomes influential at energy scales much smaller than those at which the main effects
caused by the competition between $J$ and $J_K$ start to appear, as we shall discuss later. For the time being, let us
consider the Hamiltonian (\ref{Ham-dimer}). This model was originally
studied with NRG by Jones and Varma~\cite{Jones87,Jones88,Jones89}. They found that the phase diagram
includes two stable phases. When the Kondo temperature, $T_K$, is much larger than $J$, each impurity is
Kondo screened by its conduction bath. On the contrary, when $J\gg T_K$, the two impurities lock into a singlet and
no Kondo screening is required anymore. These two stable phases were found to be separated by a critical point with
non-Fermi liquid properties~\cite{Jones89}. We notice that, since $T_K$ is a decreasing function of $U$, the
phase diagram at fixed $J/U\ll 1$ as a function of $U/\Gamma$, where $\Gamma=\Gamma(0)$ is the hybridization width defined through
\be
\Gamma(\epsilon) = \pi\,\sum_{\bk}\, V_\bk^2 \,\delta\left(\epsilon - \epsilon_\bk\right),
\label{Gamma(e)}
\ee
also includes a critical point at some $(U/\Gamma)_*$, see Fig.~\ref{dimer-phd}.
More specifically, since the Kondo temperature in units of half the conduction bandwidth behaves, at $J=0$ and for large $U$, 
as~\cite{Hewson}
\be
T_K \sim \sqrt{\frac{8\Gamma}{\pi U}}\;\mathrm{e}^{-\pi U/8\Gamma},
\label{estimate-TK}
\ee
one expects the critical point to occur approximately around  
\be
\frac{U}{\Gamma} \sim \frac{8}{\pi}\,\ln\frac{1}{J}.
\label{estimate-U/Gamma}
\ee 
In other words, perturbation theory breaks down at a finite value of the interaction within
model (\ref{Ham-dimer}), which is {\sl per se} an interesting situation uncommon in interacting Fermi systems.

The detailed properties of the critical point were later unraveled in Refs.~\cite{Affleck92PRL} and
\cite{Affleck95} by means of conformal field theory (CFT). 
The use of CFT to study impurity models relies on the fact that only a finite number of conduction electron  
scattering channels are hybridized with the impurity. This implies that, when $\Gamma(\epsilon)$, Eq.~(\ref{Gamma(e)}), 
is smooth around the chemical potential, $\epsilon=0$, on a scale larger than the Kondo temperature, 
the asymptotic low temperature/frequency behavior is similar to a  
conventional one-dimensional semi-infinite chain of non-interacting electrons, the impurity sitting at the edge. 
It is known that non-interacting electrons in one dimension can be mapped through bosonization~\cite{Sasha} 
onto a 1+1 critical field theory -- the criticality corresponding to the fermionic spectrum being gapless -- 
that is not only scale but also conformal invariant.   
This allows to fully identify and classify all critical properties: thermodynamic quantities, correlation functions and  
finite-size energy spectra~\cite{Zamo,Zamo1,Cardy1,DiFrancesco}.
In impurity models, the effective one-dimensional chain turns out to be semi-infinite, implying that 
one has actually to deal with a boundary CFT~\cite{Cardy}. Since a single impurity can not induce any gap in the 
bulk spectrum, the conformal invariance of the free electrons remains intact; the only effect of the impurity is to 
change the boundary conditions (BCs) among the conformally invariant ones. 
A crucial step to determine the allowed BCs is the so-called {\sl conformal
embedding}~\cite{DiFrancesco}, which amounts to identifying the
conformal field theories corresponding to the symmetry groups 
under which the Hamiltonian of the conduction electrons plus the impurity stays
invariant. Note that the number of gapless degrees of freedom must be conserved, which corresponds, 
within CFT, to the fact that the 
CFTs in the absence and presence of the impurity have the same total {\sl central charge}~\cite{DiFrancesco}.
A conformal embedding can be justified rigorously 
by identifying the partition function of free electrons with that obtained by combining the partition functions of the 
CFTs that emerge out of the embedding. Some simple applications of this very powerful method are given in the Appendix.  
In the most favorable cases, the proper BCs correspond to conformally invariant BCs only within one of
the different CFTs of the embedding. The next useful information
is that the conformally invariant BCs within each sector can be
obtained by the so-called {\sl fusion
hypothesis}~\cite{Cardy,affleck-1995-26}, according to which, starting
from the spectrum of a known BC, one can obtain all the others upon
{\sl fusion} with the proper scaling fields, called {\sl primary
fields}, of the CFT, see the Appendix. {\sl Fusion} is just the technical word to denote
the process in which the impurity with its quantum numbers 
dissolves into the conduction electron Fermi sea.

Let us consider for instance model (\ref{Ham-dimer}), and assume, for simplicity, that the two baths are
particle-hole invariant. As a consequence, the baths, in the absence of the impurities,
can be described by a CFT~\cite{affleck-1995-26} which includes
independent spin $SU(2)_1$ and charge isospin $SU(2)_1$ for each
bath, see Appendix, namely an overall
\[
\big(SU(2)_1^{(1)}\times SU(2)_1^{(2)}\big)_{charge}\,\times \,
\big(SU(2)_1^{(1)}\times SU(2)_1^{(2)}\big)_{spin}.
\]
The subscript in $SU(2)_k^{(a)}$ can be regarded here as the
number of copies of spin- or isospin-1/2 electrons participating to the
$SU(2)$ algebra~\cite{Sasha}, while the superscript refers to the bath.
Since the charge isospin generators
commute with the Hamiltonian even when the coupling to the impurities is switched on,
the charge sector can still be represented by two independent
isospin $SU(2)_1$ CFTs. On the other hand, only the total spin
$SU(2)$ transformations leave the Hamiltonian invariant, which
translates into an $SU(2)_2$ (two copies of electrons) CFT. As a
result the proper embedding in the spin sector
is~\cite{Affleck92PRL}, see also the Appendix, 
\[
\Bigg(SU(2)_1^{(1)}\times SU(2)_1^{(2)}\Bigg)_{spin} \rightarrow SU(2)_2 \times Z_2,
\]
where $Z_2$ denotes an Ising CFT which reflects the symmetry under
permutation of the two baths. The Ising CFT has three primary fields, the identity $I$, with dimension 0, 
$\epsilon$ (the thermal operator), with dimension 1/2, 
and $\sigma$ (the Ising order parameter), with dimension 1/16. 
Each state and operator can then be identified by the 
quantum numbers $(I_1,I_2,S,\mathrm{Ising})$ where $I_1$ and $I_2$ refer to the charge isospin of each channel, 
$S$ to the total spin and $\mathrm{Ising}$ to the $Z_2$ sector. 

Affleck and Ludwig~\cite{Affleck92PRL,Affleck95} realized that the different fixed
points found by NRG, see Fig.~\ref{dimer-phd}, correspond to the three different BCs of the Ising CFT,
namely two fixed BCs, the stable phases -- where one Ising-spin
orientation is prohibited at the boundary -- and one free BC, the
unstable fixed point -- where both orientations are allowed.
Starting from the unscreened phase, the Kondo screened phase is
obtained by fusion with the Ising primary field $\epsilon$, while the unstable critical point is obtained by fusion with the
Ising order parameter $\sigma$. Furthermore, the critical point is identified
by a finite residual entropy $S(T=0)= \frac{1}{2}\;\ln 2$, showing that part of the impurity degrees of freedom
remains unscreened.

\begin{figure}
\centerline{\includegraphics[width=8cm]{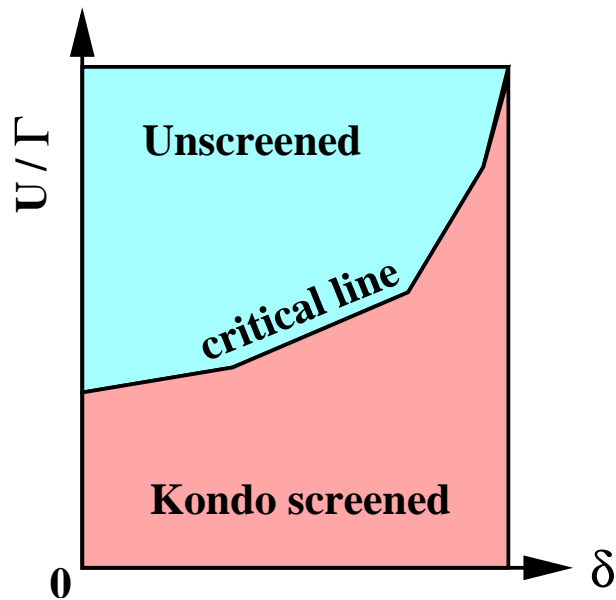}}
\caption{\label{dimerpdmu} Phase diagram of (\ref{Ham-dimer}) as function of $U/\Gamma$ and
$\delta = |\langle n\rangle -1|$, $\langle n\rangle$ being the average occupancy of each impurity, at fixed $J/U\ll 1$.}
\end{figure}

CFT also determines, by the so-called
double-fusion~\cite{affleck-1995-26}, the 
dimensions of the relevant operators, see the Appendix. It turns out that there are
three equally relevant (i.e. with dimension less than one) 
symmetry breaking perturbations which can destabilize the unstable
fixed point, identified by two asterisks in Table~\ref{k=2-fixed-content} of the Appendix. 
They all have the same dimension $1/2$ as the
invariant operator, single asterisk in Table~\ref{k=2-fixed-content} of the Appendix, which moves away from the fixed point and
corresponds to a deviation of $J$ from its fixed point value $J_*$
at fixed $T_K$, or, vice versa, a deviation of $T_K$ at fixed $J$.
The first symmetry breaking operator is an opposite spin magnetization for the
two baths, namely a local operator of the form
\be
h_{AF}\,\Big(\mathbf{S}_1 - \mathbf{S}_2\Big), 
\label{dimer-AF}
\ee
and corresponds in Table~\ref{k=2-fixed-content} of the Appendix to the operator with quantum numbers 
$(I_1,I_2,S,\mathrm{Ising})=(0,0,1,0)$. 
The second is a BCS term in the inter-bath Cooper
singlet channel:
\be
h_{SC}\,\left( d^\dagger_{1\uparrow}d^\dagger_{2\downarrow} + d^\dagger_{2\uparrow}d^\dagger_{1\downarrow}\right)
+ H.c.\,.
\label{dimer-SC}
\ee
The last perturbation is a direct
hybridization between the two baths,
\be
\sum_\sigma\, h_\perp\,d^\dagger_{1\sigma}d^\pdag_{2\sigma}\, + \, H.c.\, ,
\label{dimer-tperp}
\ee
which breaks the $O(2)$ channel symmetry. Both (\ref{dimer-SC}) and (\ref{dimer-tperp}) correspond 
in Table~\ref{k=2-fixed-content} of the Appendix to the spin-singlet operator with quantum numbers 
$(I_1,I_2,S,\mathrm{Ising})=(1/2,1/2,0,0)$. On the contrary, both a chemical potential shift that
moves away from particle hole symmetry, or a perturbation that splits the two conduction channels, do not
destabilize the critical point. Indeed, if the position of the impurity levels is modified, so that the
average number of electrons on each impurity moves away from $\langle n_a\rangle = 1$, $a=1,2$, the critical
point is still encountered, although it will shift to larger $U/\Gamma$ at fixed $J/U\ll 1$~\cite{DeLeo04f}, see
Fig.~\ref{dimerpdmu}.

\subsection{Dynamical behavior of the impurity dimer}

The instability channel (\ref{dimer-tperp}) is very important since, as we mentioned, the inter-bath as well as the inter-impurity
hybridization are always present. Therefore the relevant issue becomes whether this
hybridization completely washes out the critical behavior of the underneath
critical point, or whether a critical region still remains. In order to answer this question, it is
convenient to analyze the impurity spectral function, which is also the key ingredient of the
DMFT self-consistency procedure.

\begin{figure}
\centerline{\includegraphics[width=12cm]{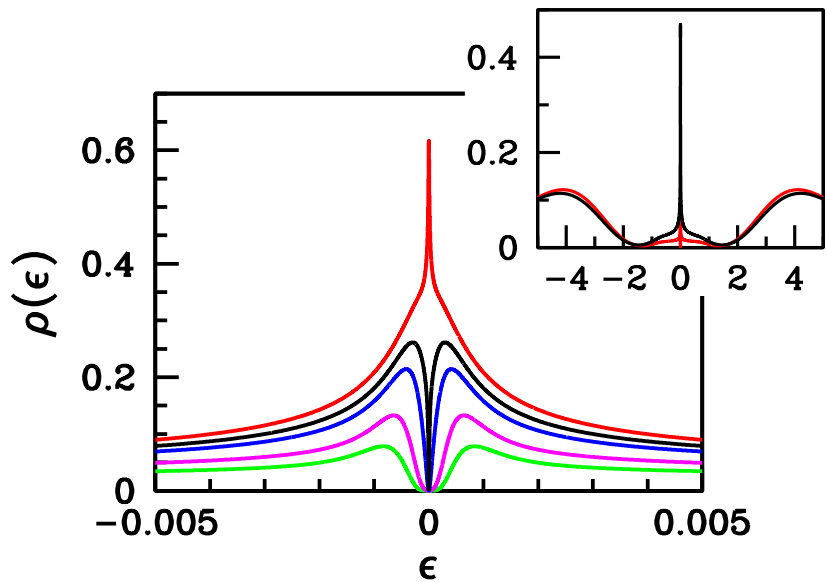}}
\caption{\label{DOSexE} Main panel: low energy behavior of the impurity DOS of the dimer model
(\ref{Ham-dimer}) with $U=8$, $J=0.00125$ and, from top to bottom, $\Gamma=0.44,~0.42,~0.4,~0.35,~0.3$, in
units of half the conduction bandwidth. 
We observe that the rough estimate of the order of magnitude of 
the critical $U/\Gamma \sim -(8/\pi)\ln J \simeq 17.3$, see Eq.~(\ref{estimate-U/Gamma}), agrees  
with the actual numerical value $(U/\Gamma)_* \sim 18.2$.  
Upper inset: the DOS behavior in the whole energy range
with the same $U$ and $J$ and with $\Gamma=0.6$, top curve, and $\Gamma=0.3$. 
The Hubbard bands are clearly visible, while the low energy parts are hardly distinguishable. 
The discretization
parameter~\cite{Krishnamurthy80i,Krishnamurthy80ii} that we used is $\Lambda = 2$.
}
\end{figure}

We start by noticing that, in spite of the fact that both Kondo screened and unscreened
phases of the Hamiltonian (\ref{Ham-dimer}) are Fermi-liquid-like in Nozi\`eres' sense~\cite{NozieresJLTP} 
(namely they correspond asymptotically to well defined limits of free-electrons scattering off a structure-less impurity potential, 
infinite in the Kondo screened phase and zero in the unscreened one),  
the dynamical properties of the impurities are completely different. Indeed,
the conduction-electron scattering $S$-matrix at the chemical potential should satisfy
\be
\mathcal{S} = 1 - 2\,\frac{\rho(0)}{\rho_0},
\label{rho.vs.S}
\ee
where $\rho(0)$ is the actual density of states (DOS) of the impurity at the chemical potential,
while $\rho_0=1/(\pi\Gamma)$ is its non interacting $U=J=0$ value. Since both stable phases are Fermi-liquid like,
it follows that the $S$ matrix is unitary, hence can be written as $\mathcal{S} = \mathrm{e}^{2i\delta}$,
where $\delta$ is the phase-shift at the chemical potential. The Kondo screened phase is identified by a phase-shift $\delta=\pi/2$,
that implies $\mathcal{S}=-1$ hence $\rho(0)=\rho_0$; the DOS at the chemical potential is unaffected
by the interaction. On the contrary, in the unscreened phase $\delta=0$, thus $\mathcal{S}=1$ and
$\rho(0)=0$. In other words, while in the Kondo screened phase the DOS is peaked at the chemical potential 
-- the conventional Kondo resonance behavior -- it vanishes in the unscreened one. Furthermore, according to
CFT, right at the critical point $\mathcal{S}=0$, namely $\rho(0)=\rho_0/2$. These results are actually
reproduced by NRG, see e.g. Ref.~\cite{DeLeo04f}. In Fig.~\ref{DOSexE}, we draw our NRG results
for the impurity DOS of the dimer model (\ref{Ham-dimer}) for $U=8$, $J=0.00125$, in units of half the
conduction bandwidth, and for various values of $\Gamma$ across the critical point $\Gamma_*$, which lies between
0.42 and 0.44. The upper inset shows that, on large scales, the DOSs in the screened and unscreened phases
are practically indistinguishable. The differences emerge at very low energies. Apart from the value of the
DOS at the chemical potential, which, as mentioned, can be anticipated by general scattering theory arguments,
other useful information can be extracted from the whole low-energy behavior. As was realized in Ref.~\cite{DeLeo04f},
the DOS is controlled by two energy scales. In the Kondo screened phase,
there is a narrow Kondo peak on top of a broad resonance. The Kondo peak shrinks as the critical point
is approached, while the width of the large resonance remains practically constant.
Indeed, right at the critical point, only the latter survives. On the contrary, inside the unscreened phase,
the Kondo peak turns into a narrow pseudo-gap within the broad resonance, leading to 
a low-energy DOS $\rho(\epsilon) \sim \epsilon^2$.
This behavior has been found to be well reproduced by the following model-DOS at low energy~\cite{DeLeo04f}:
\be
\rho_\pm(\epsilon) = \frac{\rho_0}{2}\,\Bigg(
\frac{T_+^2}{\epsilon^2+T_+^2}\pm \frac{T_-^2}{\epsilon^2+T_-^2}\Bigg),
\label{model-DOS}
\ee
where the + sign refers to the screened phase and the - one to the unscreened. $T_+$ is the width of the
broad resonance, $T_-$ the one of the narrow peak in the screened phase and the amplitude of the 
pseudo-gap in the unscreened regime. As the critical point is approached on both sides,
$T_-\sim |\Gamma -\Gamma_*|^2\to 0$~\cite{DeLeo04f}, in accordance with the CFT prediction that the relevant operator has
dimension 1/2. Away from particle-hole symmetry, $\langle n_a\rangle \not = 1$, the two stable phases are still identified
by a relative $\pi/2$-shift of $\delta$, although the unscreened phase value, $\delta_0$, is different from zero. In this case, the 
following model-DOS was found to reproduce well the NRG data~\cite{DeLeo04f}:
\be
\rho_\pm(\epsilon) = \frac{\rho_0}{2}\,\Bigg[
\frac{\displaystyle T_+^2 + \mu_{\pm}^2}{\displaystyle \left(\epsilon +\mu_{\pm}\right)^2 + T_+^2}
\pm \cos 2\delta_0\, \frac{\displaystyle T_-^2}{\epsilon^2 + T_-^2}\Bigg],
\label{model-DOS-no-ph}
\ee
with $\mu_{\pm} = \pm T_+\,\sin 2\delta_0$. This formula shows that, in the unscreened phase, the pseudo-gap
remains pinned at the chemical potential, even if, since the broad resonance shifts, the pseudo-gap fills in
of an amount proportional to the ``doping'', i.e. $|\langle n_a\rangle - 1|$.

\begin{figure}
\centerline{\includegraphics[width=12cm]{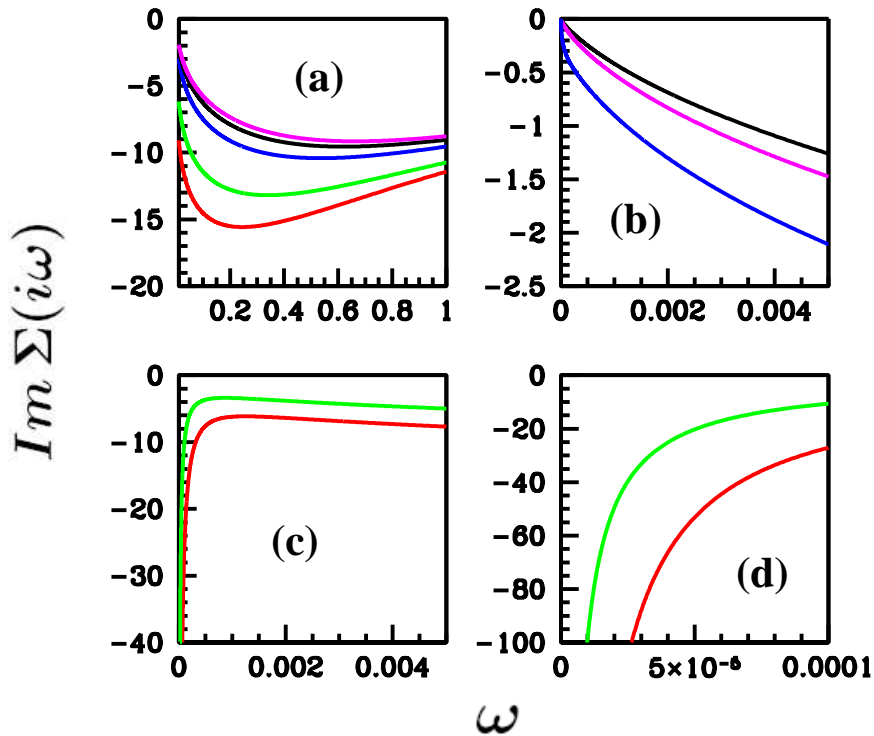}}
\caption{\label{SigmaexE} Imaginary part of $\Sigma(i\omega)$ versus $\omega$ for $U=8$
and $J=0.00125$: (a) whole frequency-range behavior for $\Gamma=0.5,~0.48,~0.44,~0.35,~0.3$, from
top to bottom; (b) low frequency behavior for the Kondo screened values $\Gamma=0.5,~0.48,~0.44$;
(c) and (d) low frequency behavior for the unscreened values $\Gamma=0.35,~0.3$. These results
were obtained with $\Lambda = 2$.}
\end{figure}

By the model DOS (\ref{model-DOS}), one can extract a model self-energy, $\Sigma(i\omega)$
in Matsubara frequencies, which, for low $\omega>0$, behaves as
\be
\Sigma_+(i\omega) \simeq - i\omega\,\Bigg(
\frac{\displaystyle \Gamma\left(T_+ + T_-\right)}{\displaystyle 2T_+ T_-} - 1\Bigg)
+ i\omega^2\, \frac{\displaystyle \Gamma\,\left(T_+-T_-\right)^2}{\displaystyle 4T_+^2 T_-^2},
\label{Sigma_+}
\ee
in the Kondo screened phase, as
\be
\Sigma_-(i\omega) \simeq -i \frac{1}{\omega}\,\frac{2\Gamma T_+ T_-}{T_+ - T_-}
- i\Gamma\,\frac{T_+ + 3T_-}{T_+ - T_-} -i\omega\,\frac{2\Gamma-T_+ + T_-}{T_+ - T_-},
\label{Sigma_-}
\ee
in the  unscreened one, and finally as
\be
\Sigma_*(i\omega) \simeq -i \Gamma -i\omega\,\frac{2\Gamma - T_+}{T_+},
\label{Sigma_*}
\ee
exactly at the critical point, $T_-=0$, or in the range $T_-\ll \omega \ll T_+$. 
The model self-energy reproduces well the actual NRG results, 
shown in Fig.~\ref{SigmaexE}. 
Only in the screened phase, panels (a) and (b) in Fig.~\ref{SigmaexE}, the self-energy has the standard perturbative
behavior,  $\Sigma(i\omega) \sim \Big(1- 1/Z\Big)\,i\omega$, with $Z$ the quasiparticle residue,
which breaks down at the critical point, where $\Sigma(i\omega)$ goes to a constant value for $\omega\to 0$, 
and even more in the unscreened regime where, as shown on different frequency ranges in panels (a), (c) and (d) of 
Fig.~\ref{SigmaexE}, $\Sigma(i\omega)$ diverges as $\omega\to 0$.

\begin{figure}
\centerline{\includegraphics[width=12cm]{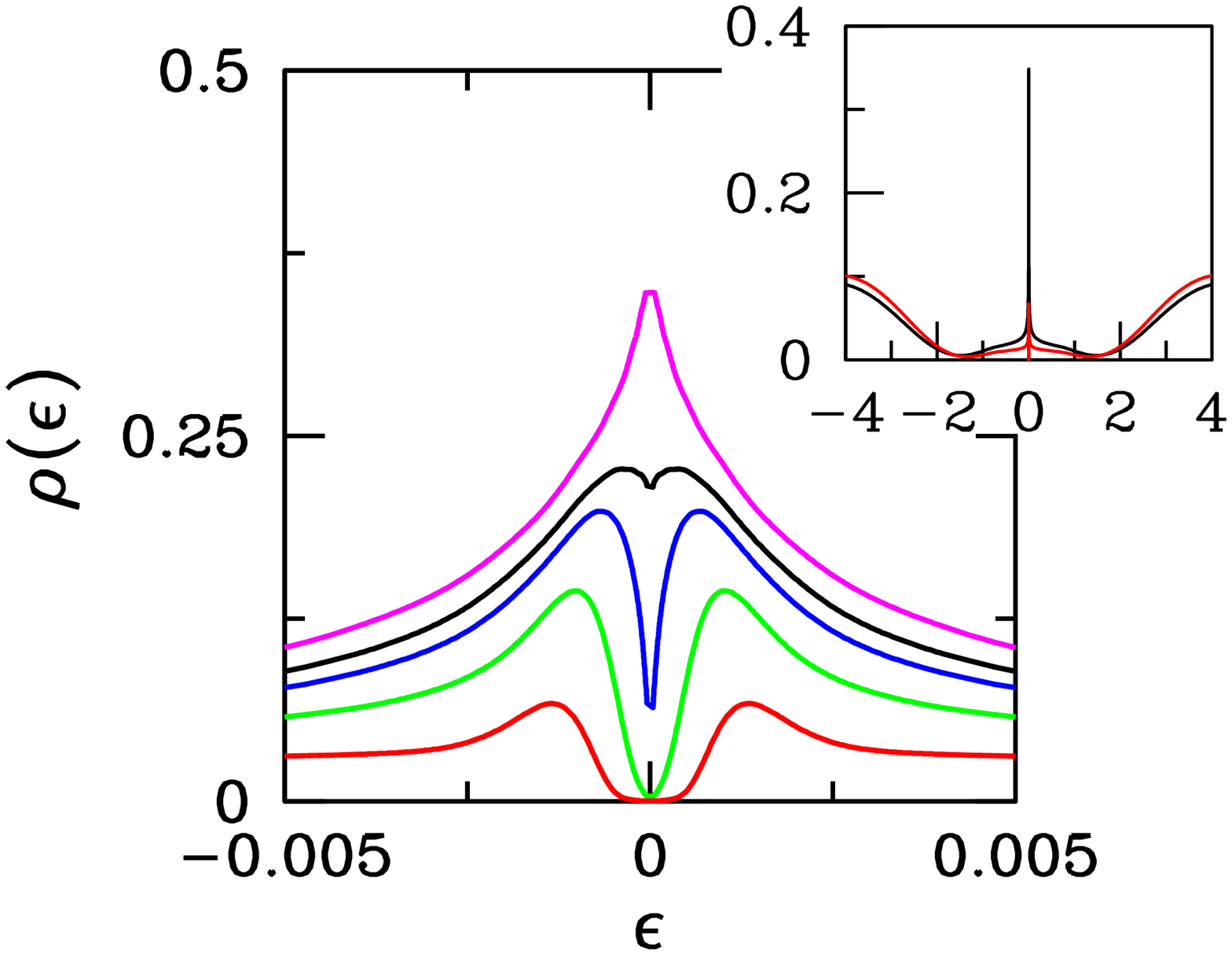}}
\caption{\label{DOStperp} Main panel: low energy behavior of the impurity DOS of the model
(\ref{AIM-dimer}) with $U=8$, $t_\perp=0.05$ and, from top to bottom, $\Gamma=0.5,~0.47,~0.45,~0.4,~0.3$, in
units of half the conduction bandwidth. Upper inset: the DOS behavior in the whole energy range
with the same $U$ and $J$ and with $\Gamma=0.5$, top curve, and $\Gamma=0.2$. These results
were obtained with $\Lambda = 2$.}
\end{figure}

Let us now consider the original dimer model (\ref{AIM-dimer}). In order not to deal with too many
Hamiltonian parameters, we consider the case in which the inter-bath hybridization is zero, $t_{\perp\, \bk}=0$,
and take $U=8$, as before, and $t_\perp=0.05$, such that $J=4t_\perp^2/U=0.00125$, the same value used previously.
In this model $t_\perp$ has a double role: on one side it generates a spin-exchange able to drive the model across
the critical point, but at the same time it also breaks the relevant $O(2)$ channel symmetry thus making 
the critical point inaccessible. Indeed, as shown by the behavior of the impurity DOS in Fig.~\ref{DOStperp},
a crossover now joins the Kondo screened phase and the unscreened one. There are still quite distinct
DOSs deep inside the screened and unscreened phases, the former characterized by a Kondo resonance, the latter by a
pseudo-gap. However, the transition from the two limiting behaviors is now just a crossover, although quite sharp.

\begin{figure}
\centerline{\includegraphics[width=12cm]{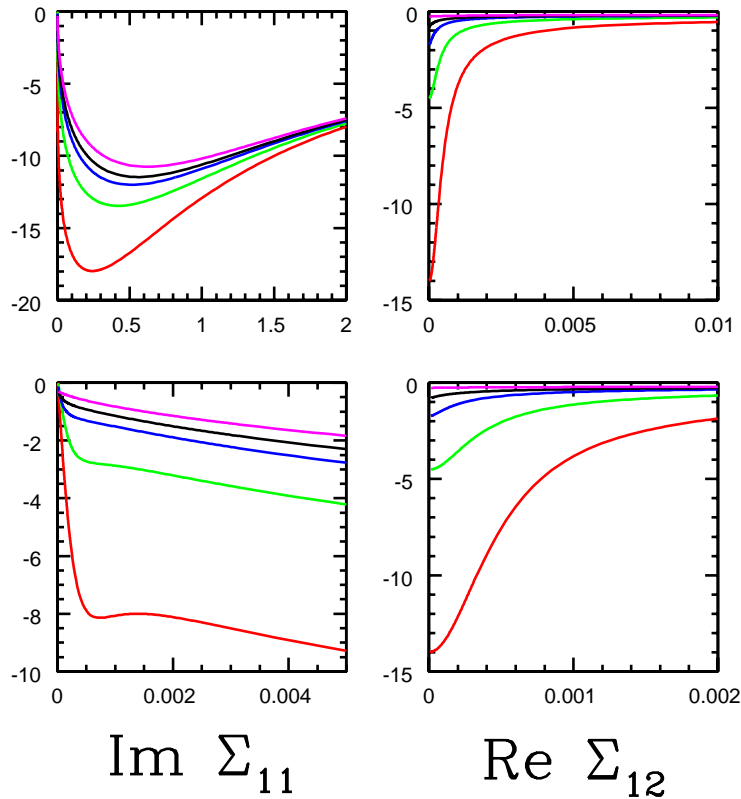}}
\caption{\label{Sigma} $y$-axis: Imaginary part of $\Sigma_{11}(i\omega)$, left panels, and real part of
$\Sigma_{12}(i\omega)$, right panels, versus $\omega$ ($x$-axis) for $U=8$, $t_\perp=0.05$ and, from top to bottom,
$\Gamma=0.5,~0.47,~0.45,~0.4,~0.3$ and $\Lambda = 2$. In the top figures the whole frequency range
is showed, while in the bottom ones only the very-low frequency behavior.}
\end{figure}

More interesting are the changes that intervene in the impurity self-energy with respect to the model (\ref{Ham-dimer}).
In presence of $t_\perp$, the self-energy, besides diagonal elements, $\Sigma_{11}(i\omega) = \Sigma_{22}(i\omega)$, which
are imaginary, also acquires off-diagonal components, $\Sigma_{12}(i\omega) = \Sigma_{21}(i\omega)^*$, which
turn out to be purely real. In Fig.~\ref{Sigma}, we plot $\mathcal{I}m\, \Sigma_{11}(i\omega)$
and $\mathcal{R}e\, \Sigma_{12}(i\omega)$ versus $\omega$ for the same values of $\Gamma$'s as in
Fig.~\ref{DOStperp}, on two different energy ranges. We notice that the gross features of $\Sigma_{11}(i\omega)$
remain intact, namely, as the model moves from the screened regime towards the unscreened one, the diagonal self-energy
increases quite fast in absolute value. However, a linearly-vanishing ``Fermi-liquid'' behavior is eventually recovered
at very low energies, as shown in the left-bottom panel of Fig.~\ref{Sigma}. In fact, the model in the presence of $t_\perp$
does not need to develop a singular self-energy anymore to open up a pseudo-gap at the chemical potential
as in model (\ref{Ham-dimer}). It is the off-diagonal self-energy, $\Sigma_{12}$, that accomplishes the job in this
case. Indeed, as shown in the right panels of Fig.~\ref{Sigma}, $\Sigma_{12}$ becomes so large at low energy to open up
an appreciable hybridization gap, not explainable by the tiny value of $t_\perp$ as compared to the hybridization width
$\Gamma$. This result could be justified simply by stating that the strong repulsion $U$ enhances the \underline{effective} 
strength of $t_\perp$. However, the previous results on the dimer-model (\ref{Ham-dimer}) and the strong energy-dependence 
of the self-energy, see Fig.~\ref{Sigma}, suggest that this anomalous behavior reflects rather the properties of the avoided 
critical point which exists in the presence of $J$ at $t_\perp=0$. In other words, we believe that our results 
testify that a sizable critical region still exits and largely explains the physical behavior.

Obviously, as in any other case
of avoided criticality, the width of the critical region depends on the actual value of $t_\perp$ with respect to the other
parameters $U$ and $\Gamma$. Since both $t_\perp$ and $\Gamma$ are self-consistently
determined within DMFT as function of $U$ and the bare
bandwidth, we cannot establish with certainty what might happen in a DMFT simulation
of a Hubbard model using a dimer as a representative
cluster. However, because the exchange $J$ derives from high-energy processes and survives even inside the Mott insulator, while the
coherent hopping dies out at the Mott transition, we believe that a sizable critical region should exist in the effective
impurity-cluster model and plays an influential role in determining the bulk properties after the DMFT
self-consistency.

\section{The impurity trimer}
\label{The impurity trimer}

\begin{figure}
\centerline{\includegraphics[width=10cm]{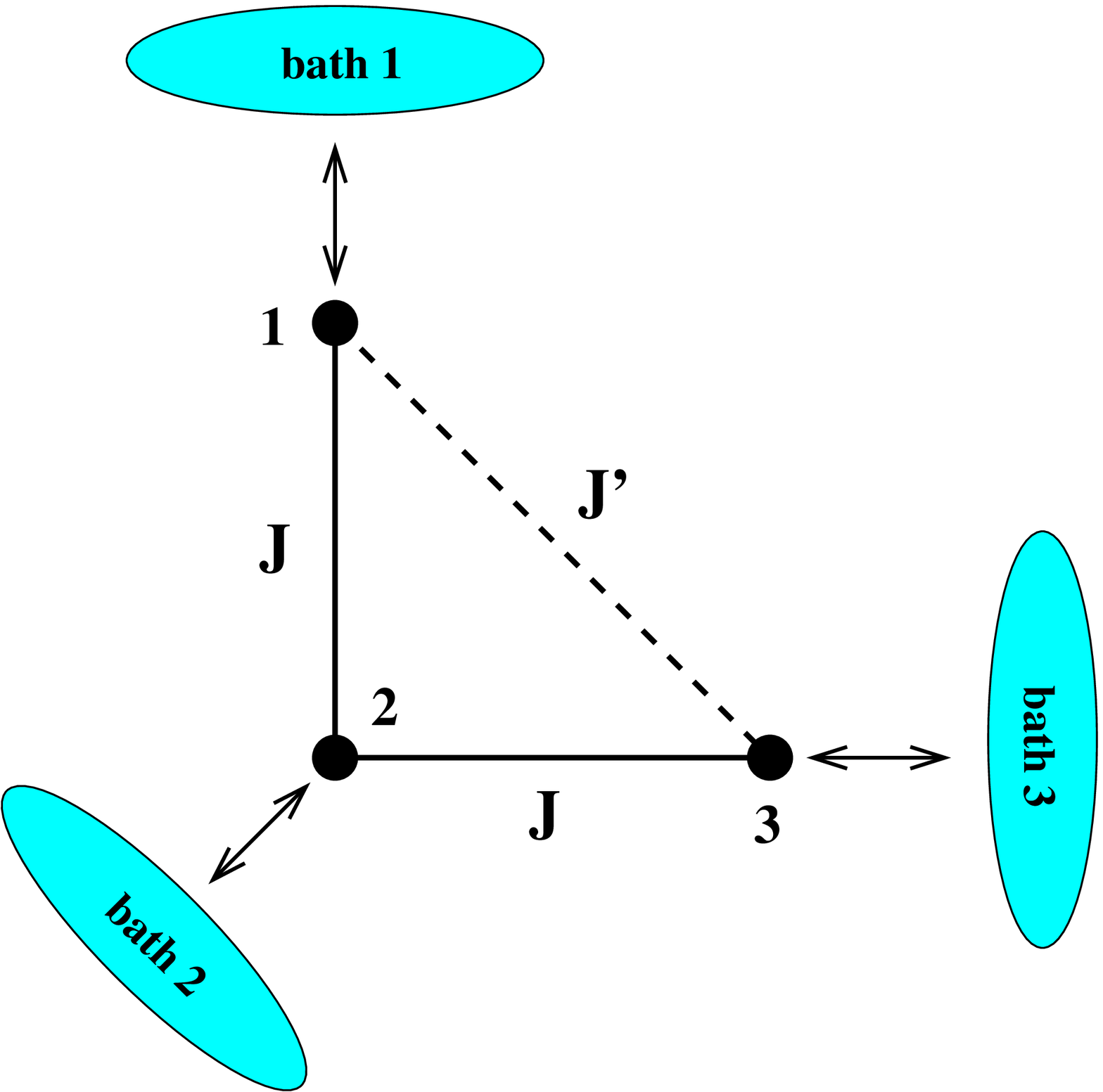}}
\caption{\label{Fig-trimer} The impurity trimer with the Hamiltonian
(\ref{Ham-trimer})}
\end{figure}

An important lesson of the impurity dimer was that, in order to identify in all details the properties of the
critical region, it is more convenient to study a model in which the impurities are coupled together by an
antiferromagnetic exchange rather than by a hopping as would be the case in reality. In the end, we will
discover that, just like in the dimer example, the hopping is a relevant perturbation, and yet a critical
region survives.
Therefore, let us consider the next simple cluster, which is
the impurity trimer drawn in Fig.~\ref{Fig-trimer} with the Hamiltonian
\be
\mathcal{H} = \sum_{a=1}^3\,\mathcal{H}_a^K\, +
J\,\left(\mathbf{S}_1+\mathbf{S}_3\right)\cdot \mathbf{S}_2
+ J'\,\mathbf{S}_1\cdot \mathbf{S}_3,
\label{Ham-trimer}
\ee
where $\mathcal{H}_a^K$ has been defined in (\ref{Ham-dimer}). This model describes
three spin-1/2 impurities, coupled together by antiferromagnetic $J$ and $J'$ and each
of them hybridized to a conduction bath. As before, we assume that the baths are degenerate
and particle-hole invariant. This model, although simplified by the absence of any inter-impurity
hopping, is much more complicated than the dimer model (\ref{Ham-dimer}). Therefore, in order to
unravel its phase diagram, we need to combine the NRG analysis, which provides the low-energy
spectra of the various fixed points, with CFT, which allows to identify each fixed point with
a particular boundary CFT, whose properties can be determined exactly. For this reason, we cannot
start our analysis before introducing some CFT preliminaries.

\subsection{CFT preliminaries for the trimer}
\label{CFT preliminaries for the trimer}

As in the dimer example, also in the trimer model
(\ref{Ham-trimer}) at particle-hole symmetry the charge degrees of freedom can be described by
three independent isospin $SU(2)_1^{(a)}$ CFTs, $a=1,2,3$. For
the spin degrees of freedom, the expression of 
the inter-impurity exchange suggests naturally that we must first 
couple the spin sectors of baths 1 and 3 into an overall
$SU(2)_2$ via the embedding
\be
SU(2)_1^{(1)}\times SU(2)_1^{(3)}
\rightarrow SU(2)_2^{(1-3)} \times Z_2,
\ee
and finally couple the $SU(2)_2$ to the bath 2 into an $SU(2)_3$, according to
\be
SU(2)_2^{(1-3)} \times SU(2)_1^{(2)} \rightarrow SU(2)_3 \times
\left({\rm TIM}\right),
\ee
where TIM stands for the tricritical Ising
model CFT with central charge $c=7/10$. 
It describes for instance the tricritical point of the 
two-dimensional Blume-Capel model which involves an Ising spin variable
and a vacancy variable indicating if the site is empty
or occupied~\cite{Blume,Capel}.
The above conformal embedding can be rigorously justified by the {\sl character
decomposition}~\cite{DiFrancesco}, although we do not give here the details of this lengthy  
and involved construction.  

The primary fields of an $SU(2)_k$ as well as of the Ising CFTs, together with their fusion, i.e. multiplicative, rules, 
are discussed in the Appendix. For what concerns the TIM, it contains 
six primary fields, the identity $I$, with dimension 0,
the thermal energy operator $\epsilon$, with dimension 1/10,
the energy density of annealed vacancies $t$, with dimension 3/5, $\epsilon^{''}$,
with dimension 3/2, the magnetization $\sigma$, with dimension 3/80, 
and the subleading magnetization operator $\sigma^{'}$, with
dimension 7/16~\cite{DiFrancesco}. Their fusion rules can be found for instance in
Ref.~\cite{DiFrancesco}, pg.~224.

As we previously mentioned, the possible conformally invariant
boundary conditions can be classified by means of the fusion
hypothesis~\cite{Affleck:1990by,Affleck:1990iv,Affleck95}. Namely,
starting from the spectrum of a simple BC, one can obtain the spectra of other allowed BCs upon
fusion with primary fields of the CFTs. By comparing the low-energy spectra determined in this way with those obtained by NRG,
one can identify and characterize all fixed points of the model.

\subsection{Fixed points in the trimer phase diagram}
\label{Fixed points in the trimer phase diagram}

\begin{figure}
\centerline{\includegraphics[width=12cm]{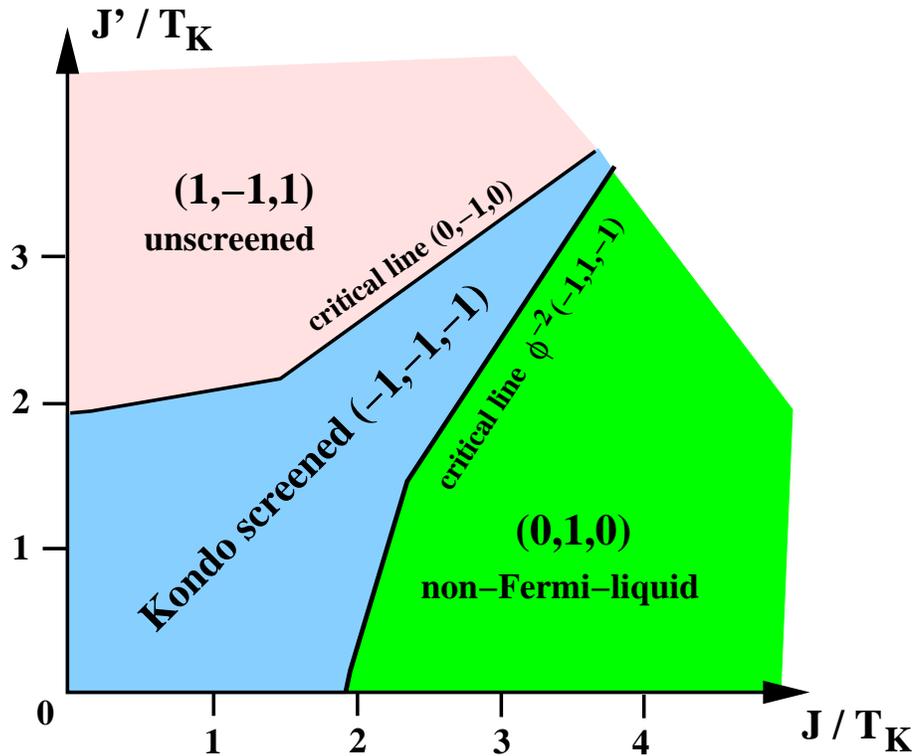}}
\caption{Phase diagram of the trimer model (\ref{Ham-trimer}). All different phases are discussed in the text, and their 
properties briefly summarized in Table~\ref{Table trimer}. 
\label{trimer-phd}}
\end{figure}

In Fig.~\ref{trimer-phd}, we draw the phase diagram of the trimer
as obtained by NRG~\cite{Wilson75,Krishnamurthy80i,Krishnamurthy80ii}.
In order to have a classification
scheme which works equally well for Fermi-liquid and
non-Fermi-liquid phases, the fixed points are identified through
the zero-frequency values of the scattering $S$-matrices of the baths,
$(S_1,S_2,S_3)$, which can be obtained by
CFT~\cite{Affleck:1991tk,Affleck93} through the modular
$S$-matrix~\cite{DiFrancesco}. Note that, through Eq.~(\ref{rho.vs.S}),
the values of the scattering $S$-matrices give direct access to the values of the DOS at the chemical potential
of each impurity. We just recall that $S_a=-1$ means that the impurity-$a$ DOS has a Kondo resonance,
$S_a=1$ that it has a pseudo-gap, $\rho_a(\epsilon)\sim \epsilon^2$, while any intermediate value implies a
non-Fermi liquid behavior. 
The physical properties of the different phases are furthermore summarized in Table~\ref{Table trimer}.

\begin{table}[htb]
\caption{Summary of the main physical properties of the different phases in Fig.~\ref{trimer-phd}, 
including the behavior of the zero-frequency DOSs for the three impurities, $\rho_i(0)$ with $i=1,2,3$, 
with respect to the non-interacting value $\rho_0$, and the dimension of the relevant 
symmetry-breaking single-particle operators. ``NOT'' means that the operator is not relevant, i.e. has dimension 
not smaller than one. $\phi$ is the golden ratio.
\label{Table trimer}}
\begin{indented}
\item[] \begin{tabular}{@{}llllll}
\br
~~~ & $(0,1,0)$ & $\phi^{-2}\,(-1,1,-1)$ & $(-1,-1,-1)$ & $(0,-1,0)$ & $(1,-1,1)$ \\
\mr
$\rho_1(0)/\rho_0=\rho_3(0)/\rho_0$ & $1/2$ & $\left(1+\phi^{-2}\right)/2$ & $1$ & $1/2$ & 0 \\
$\rho_2(0)/\rho_0$ & 0 & $\left(1-\phi^{-2}\right)/2$ & $1$ & $1$ & 1 \\ 
$\mathbf{S}_1 + \mathbf{S}_3 - 2\mathbf{S}_2$ & 1/2 & 2/5 & NOT & NOT & NOT \\
$\mathbf{S}_1 - \mathbf{S}_3$ & NOT & NOT & NOT & 1/2 & NOT \\
$d^\dagger_1 d^\pdag_3$$^a$ & 1/2 & 3/5 & NOT & 1/2 & NOT \\
$d^\dagger_1 d^\dagger_3$$^a$ & 1/2 & 3/5 & NOT & 1/2 & NOT \\
$\left(d^\dagger_1+d^\dagger_3\right)d^\pdag_2$$^a$ & NOT$^{b}$ & 3/5 & NOT & NOT$^{b}$ & NOT \\
$\left(d^\dagger_1+d^\dagger_3\right)d^\dagger_2$$^a$ & NOT$^{b}$ & 3/5 & NOT & NOT$^{b}$ & NOT \\
\br
\end{tabular}
\item[]$^a$ These particle-hole and particle-particle operators are spin-singlets.
\item[]$^b$ These operators are not relevant in the sense that their dimension is not smaller than one. However, they 
do generate in perturbation theory one of the truly relevant perturbations, with dimension smaller than one, so that 
in reality they are relevant too.  
\end{indented}
\end{table}

Let us now present briefly the features of each fixed point. 
\\[1.5ex]
\noindent \textsl{\bf 1.\quad $\mathbf{(S_1,S_2,S_3)=(-1,-1,-1)}$}
\\[1.5ex]
This fixed point, that describes a conventional perfectly Kondo-screened
phase, will be used as the ancestor BC which, upon fusion with primary fields,
will provide all other BCs.
It is quite obvious that this phase exists and extends in a whole region around the
origin $J=J'=0$ in Fig.~\ref{trimer-phd}. Indeed, when $J=J'=0$,
each impurity is independently Kondo screened by its own
conduction bath and this perfect screening cannot be affected by
finite $J$ and $J'$ much smaller than the Kondo temperature. It is
far less obvious that this fixed point remains stable for
large $J\simeq J'$. When $J'=J\gg T_K$, the impurities lock into
two degenerate S=1/2 configurations. In the first, sites 1 and 3
are coupled into a triplet which in turn is coupled with site 2
into an overall spin-1/2 configuration. Since this is even by
interchanging 1 with 3, we denote it as $|e\rangle$. The other
configuration, which we denote as $|o\rangle$ as it is odd under
$1\leftrightarrow 3$, corresponds to sites 1 and 3 coupled into a
singlet, leaving behind the free spin-1/2 moment of site 2. The
Kondo exchange projected onto this subspace reads
\bea
&&
\frac{J_K}{3}\; |e\rangle\langle e |\; \mathbf{S}\cdot
\big(2\mathbf{J}_1(0) - \mathbf{J}_2(0) + 2\mathbf{J}_3(0)\big) \nonumber \\
&& + J_K\;
|o\rangle\langle o |\; \mathbf{S}\cdot \mathbf{J}_2(0)
\nonumber \\
&& - \frac{J_K}{\sqrt{3}}\,\Big( |e\rangle\langle o | + |o\rangle\langle e |\Big)\,
\mathbf{S}\cdot \big(\mathbf{J}_1(0) - \mathbf{J}_3(0)\big),
\label{J'= J>>0}
\eea
where $\mathbf{S}$ describes the effective S=1/2 of the trimer, while $\mathbf{J}_a(0)$
is the spin density of bath $a=1,2,3$ at the impurity site, assumed to be the origin.
All the above screening channels flow to strong coupling within a simple one-loop calculation.
Since it can be readily shown that the impurity can be perfectly screened, both in the spin
and in the even-odd channels, one has to conclude that the whole line $J=J'$ at finite $J_K$
corresponds to the Kondo screened fixed point $(-1,-1,-1)$, as in Fig.~\ref{trimer-phd}.
A small deviation from $J=J'$ is an irrelevant perturbation that splits the degeneracy between
$|e\rangle$ and $|o\rangle$. Only a finite deviation eventually destabilizes this fixed point, the faster
the smaller $J_K$.
\\[1.5ex]
\noindent \textsl{\bf 2. \quad $\mathbf{(S_1,S_2,S_3)=(0,1,0)}$}
\\[1.5ex]
This fixed point occurs for $J\gg T_K,J'$, see
Fig.~\ref{trimer-phd}. The NRG spectrum is compatible with that 
obtained by fusing the $(-1,-1,-1)$ fixed point with the field $\sigma^{'}$
of the TIM. It is not difficult to realize that this fixed point
is equivalent to the non-Fermi liquid phase of the S=1/2
two-channel Kondo model~\cite{Affleck:1990by,Affleck:1990iv}.
Indeed if $J'=0$ and $J$ is large, the trimer locks into the
S=1/2 configuration which we previously denoted as $|e\rangle$, to
indicate the even parity upon $1\leftrightarrow 3$. The Kondo
exchange projected onto this configuration is, see Eq.~(\ref{J'=
J>>0}),
\be
\mathbf{S}\cdot \sum_{a=1}^3 \, J_K^{(a)}\,
\mathbf{J}_a(0) = \frac{J_K}{3}\, \mathbf{S}\cdot \big(2\mathbf{J}_1(0)
- \mathbf{J}_2(0) + 2\mathbf{J}_3(0)\big).
\label{J'=0 J>>0}
\ee
Hence, while baths 1 and 3 are still antiferromagnetically coupled, the
coupling with bath 2 turns effectively ferromagnetic. The ordinary
one-loop renormalization group calculation would predict that the
Kondo exchanges $J_K^{(1)}=J_K^{(3)}>0$ flow towards strong
coupling, while $J_K^{(2)}<0$ flows towards zero. This suggests
that a model with $J_K^{(1)}=J_K^{(3)} \gg -J_K^{(2)} >0$ should
behave asymptotically as (\ref{J'=0 J>>0}). If $J_K^{(2)}=0$ this
is just the two-channel spin-1/2 impurity
model~\cite{Affleck:1990by,Affleck:1990iv}, which is non
Fermi-liquid with $S$-matrices
$S_1=S_3=0$~\cite{Affleck:1991tk,Affleck93}. It is easy to show 
that the small ferromagnetic $J_K^{(2)}$ transforms into an
antiferromagnetic exchange of the form $\mathbf{J}_2(0)\cdot\left(\mathbf{J}_1(0)+\mathbf{J}_3(0)\right)$, 
which is irrelevant. Consequently, we expect that
this phase should remain non-Fermi liquid and identified by the
$S$-matrices $(S_1,S_2,S_3)=(0,1,0)$, as indeed confirmed by CFT.
In addition, through the modular $S$-matrix, one can show that the
zero-temperature entropy $S(0) = 1/2 \, \ln 2$ is finite and
coincides with that of the S=1/2 two channel Kondo model.
Since $\sigma^{'}\times \sigma^{'} = I + \epsilon^{''}$, with the
latter having dimension $3/2>1$, this fixed point is stable to
symmetry-preserving perturbations.
Yet, there are several
symmetry-breaking relevant perturbations of dimension 1/2. One of
them corresponds to the staggered magnetization
\be
\boldsymbol{J}_1- 2\boldsymbol{J}_2 +
\boldsymbol{J}_3.
\label{trimer-Ms}
\ee
All the other
relevant operators break the degeneracy between bath 1 and 3,
as for instance the spin-singlet operator
\be
\boldsymbol{J}_2 \cdot \big(\boldsymbol{J}_1 -
\boldsymbol{J}_3\big),
\label{trimer-s2(s1-s3)}
\ee
and the direct hopping or singlet-pairing between baths 1 and 3, both known to be relevant perturbations at the overscreened
non-Fermi-liquid fixed point~\cite{Affleck92PRB}.  
Note that, although the hopping/pairing operators between
baths 1 and 2 as well as 3 and 2 have dimension one,   
they do induce indirectly a symmetry-breaking coupling among baths 1 and 3 of dimension 1/2, 
hence they are effectively relevant, specifically marginally relevant.
This phase with $S$-matrices $(S_1,S_2,S_3)=(0,1,0)$ 
extends at finite $J'$ just because $J'$ does not generate
any symmetry-breaking relevant perturbation. 

The approach to the fixed point is controlled by two leading
irrelevant operators of dimension 3/2: $\epsilon^{''}$ and the
scalar product of the staggered magnetization (\ref{trimer-Ms})
with the first spin descendant. Similarly to the overscreened
two-channel Kondo model~\cite{Affleck:1990by,Affleck:1990iv}, these
operators produce logarithmic singularities in the impurity
contribution to the specific heat coefficient and to the magnetic
susceptibility, $C_{imp}/T \sim \chi_{imp} \sim \ln(1/T)$.
\\[1.5ex]
\noindent \textsl{\bf 3.\quad $\mathbf{(S_1,S_2,S_3)= \phi^{-2}\, (-1,1,-1)}$}
\\[1.5ex]
Since the Kondo screened phase, $(-1,-1,-1)$ and the non-Fermi
liquid one, $(0,1,0)$, are essentially different, it is clear
that an unstable critical line separates the two, see
Fig.~\ref{trimer-phd}. We find that the NRG spectrum can be
reproduced by fusing the $(-1,-1,-1)$ fixed point with $\epsilon$ of the TIM.
The $S$-matrices are $\phi^{-2}\,(-1,1,-1)$ and the residual
entropy is $S(0) = \ln \phi $, where $\phi = (1+\sqrt{5})/2$ is
the golden ratio. Since $\epsilon\times \epsilon = I + t$, the
operator which moves away from the critical line has dimension
3/5. The most
relevant symmetry breaking operator is still the staggered
magnetization (\ref{trimer-Ms}), which has now dimension 2/5. Once
more, the approach to this fixed point is controlled by the scalar
product of the staggered magnetization with the first Kac-Moody
descendant of the $SU(2)_3$ CFT, which has dimension $1+2/5$.
Analogously to the multichannel
Kondo~\cite{Affleck:1990by,Affleck:1990iv}, this operator produces
impurity contributions to the specific-heat coefficient and
magnetic susceptibility that diverge like $T^{-1/5}$.

The spin-singlet operator (\ref{trimer-s2(s1-s3)}) is also
relevant, although with a larger dimension 3/5. In addition, there
is a new class of dimension-3/5 operators which correspond to
coupling into a spin-singlet two particles, or one hole and one
particle, belonging to bath 2 and either bath 1 or 3.
Finally, this critical line is stable towards moving away from particle-hole symmetry, as it was the case
in the dimer.
\\[1.5ex]
\noindent \textsl {\bf 4.\quad $\mathbf{(S_1,S_2,S_3)=(1,-1,1)}$ and $\mathbf{(S_1,S_2,S_3)=(0,-1,0)}$}
\\[1.5ex]
These two fixed points occur when $J'>J$ is larger or comparable
with the Kondo temperature. They have a very simple explanation.
Indeed, when $J=0$, site 2 is only coupled to bath 2 with a Kondo
exchange, leading to a full screening, i.e. $S_2=-1$. Sites 1 and
3 plus their own baths realize a two-impurity Kondo
model which, as discussed before, has two stable regimes. One is
Kondo screened, $S_1=S_3=-1$, for $J'\ll T_K$,
$(S_1,S_2,S_3)=(-1,-1,-1)$ in Fig.~\ref{trimer-phd}, and the other
unscreened for $J'\gg T_K$, $(S_1,S_2,S_3)=(1,-1,1)$ in
Fig.~\ref{trimer-phd}. These two regimes are stable
towards switching on a small $J\ll J'$. When $J=0$, we also know
that an unstable fixed point at $J'=J'_*\sim T_K$ separates these
two stable phases, which is identified by $S_1=S_3=0$, hence the
label (0,-1,0) in Fig.~\ref{trimer-phd}. Since site 2 is tightly
bound into a singlet state with bath 2, a finite but small $J\ll
T_K$ will simply generate a ferromagnetic exchange of order
$-J^2/T_K$ by virtually exciting the singlet state. The net effect
is that the unstable fixed point at $J=0$ is just the endpoint of
another critical line which, for $J\ll T_K$, moves to larger
values of $J'$. From the CFT viewpoint, the $(1,-1,1)$ and
$(0,-1,0)$ fixed points can be obtained by fusing with $\epsilon^{''}$ of the TIM
or $\sigma_I$ of the Ising CFT, respectively. The properties of
the unstable $(0,-1,0)$ critical line are the same as those of the
dimer critical point. In particular there is a relevant operator
in the singlet Cooper channel that now involves pairing among
baths 1 and 3, as well as an equally relevant operator which
corresponds to an opposite magnetization of bath 1 and 3, i.e.
$\mathbf{J}_1 - \mathbf{J}_3$.

\subsection{Dynamical properties of the trimer}

In the $J$-$J'$ parameter-space of the trimer model (\ref{Ham-trimer}), the
$J'=0$ line is actually pertinent to a Hubbard model on a bipartite lattice and with nearest neighbor hopping
simulated within cluster-DMFT. In this case, the phase diagram is qualitatively similar to that of the
dimer model, see Fig.~\ref{dimerpdmu}, apart from the fact that the unscreened phase corresponds now to the
$(S_1,S_2,S_3)=(0,1,0)$ non-Fermi liquid phase, while the critical line represents the
$(S_1,S_2,S_3)= \phi^{-2}\, (-1,1,-1)$ unstable fixed point.

\begin{figure}
\centerline{\includegraphics[width=12cm]{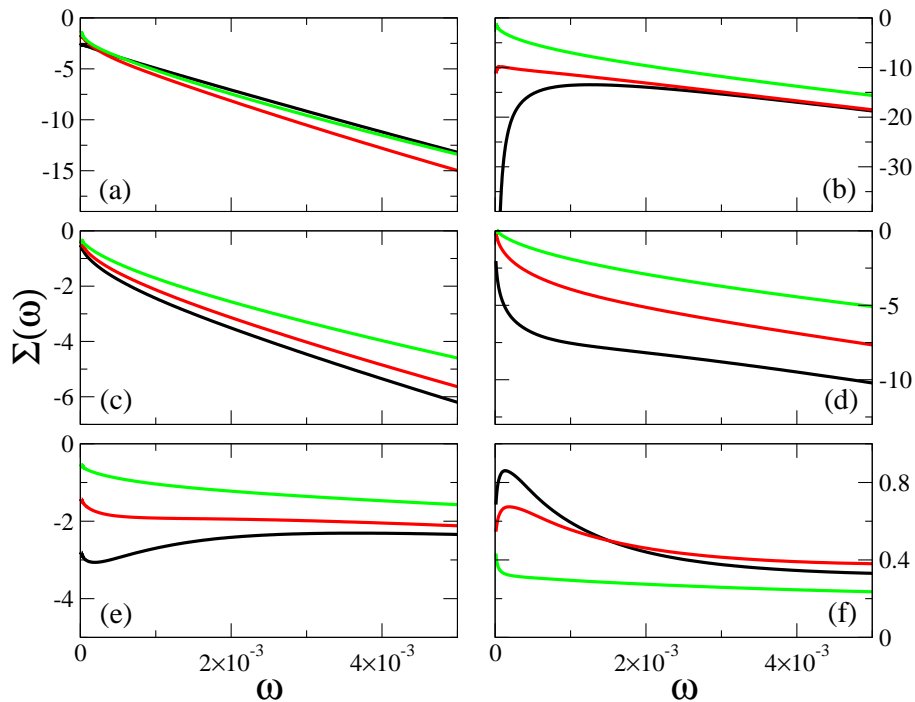}}
\caption{Non-vanishing self-energies of the trimer. The three curves correspond to $\Gamma=0.5$ (green), 0.44 (red) and 0.4 (black). Panels (a) $\Ima \Sigma_{11}(i\omega) = \Ima \Sigma_{33}(i\omega)$
and (b) $\Ima \Sigma_{22}(i\omega)$, for the case of 3 impurities coupled by the spin exchange $J$.
Panels (c) $\Ima \Sigma_{11}(i\omega) = \Ima \Sigma_{33}(i\omega)$, (d) $\Ima \Sigma_{22}(i\omega)$,
(e) $\Rea \Sigma_{12}(i\omega) = \Rea \Sigma_{32}(i\omega)$ and (f) $\Ima \Sigma_{13}(i\omega)$, for the case
of impurities coupled by a single-particle hopping $t_\perp$, leading to the same value of $J$ as before.
We used a discretization parameter $\Lambda = 3.0$.
\label{Sigmatrimer}}
\end{figure}

Both the critical line and the non-Fermi liquid phase are unstable to a nearest-neighbor hopping. 
If, instead of three impurities coupled by $J$, we consider three impurities coupled by a hopping,
the transition from the Kondo screened phase to the non-Fermi liquid one transforms into a crossover
between Fermi-liquid regimes, exactly like in the dimer. In Fig.~\ref{Sigmatrimer}, we draw the self-energies
in Matsubara frequencies of the 3 impurities coupled by the spin-exchange $J$ (panels (a) and (b)),
as well as of the impurities coupled by a hopping $t_\perp$ along the bonds 1-2 and 2-3 (panels (c) to (f)). The Hamiltonian parameters
are $U=8$, $J=0.00125$ and $t_\perp=0.05$, like in Figs.~\ref{SigmaexE} and \ref{Sigma}, while $\Gamma= 0.4,~0.44,~0.5$.
In the absence of $t_\perp$, the self-energies, which are diagonal and imaginary, behave as $\Ima \Sigma_{aa}(i\omega) \sim -\omega$
in the Kondo screened phase, $\Gamma=0.5$, and tend to a constant at the fixed point, $\Gamma\simeq 0.44$. In the unscreened
phase, $\Gamma=0.4$, $\Ima \Sigma_{11}(i\omega) = \Ima \Sigma_{33}(i\omega) \to \mathrm{const.}$
while $\Ima \Sigma_{22}(i\omega) \sim -1/\omega$, in accordance with the values of the scattering $S$-matrices.
Similarly to the dimer, when the impurities are instead coupled
by a single-particle hopping, a Fermi-liquid behavior $\Ima \Sigma_{aa}(i\omega) \sim -\omega$ is eventually recovered at very low
energy, see panels (c) and (d), although the DOS at site 2 may still display a pseudo-gaped behavior due to the
large value of $\Rea \Sigma_{12}(i\omega) = \Rea \Sigma_{32}(i\omega)$.

In conclusion, the dynamical behavior with $J$ or in the presence of $t_\perp$ is qualitatively
similar to the 2-impurity cluster.

\section{The impurity tetramer}

\begin{figure}
\centerline{\includegraphics[width=10cm]{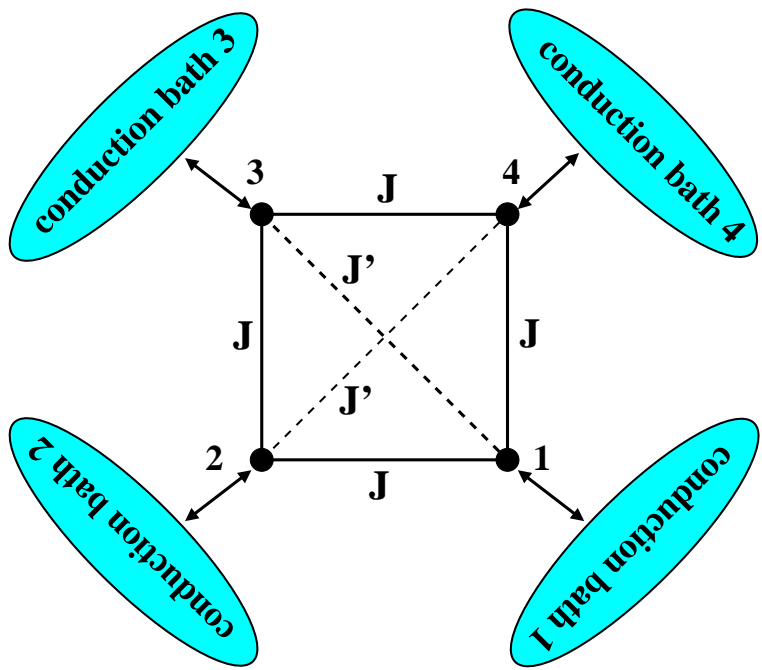}}
\caption{\label{Fig-tetramer} The 4-impurity cluster}
\end{figure}

Let us move finally to the last type of cluster investigated, the tetramer drawn in Fig.~\ref{Fig-tetramer}
with the Hamiltonian
\bea
\mathcal{H} &=& \sum_{a=1}^4\,\mathcal{H}_a^K\, +
J\,\left(\mathbf{S}_1+\mathbf{S}_3\right)\cdot \left(\mathbf{S}_2 + \mathbf{S}_4\right)\nonumber \\
&& + J'\,\left(\mathbf{S}_1\cdot \mathbf{S}_3 + \mathbf{S}_2\cdot \mathbf{S}_4\right).
\label{Ham-tetramer}
\eea
This model now describes four
spin-1/2 impurities, coupled together by nearest $J$, and next-nearest
neighbor $J'$, antiferromagnetic exchanges. In addition, each spin is Kondo-coupled
to a conduction bath by $J_K>0$. The four baths are once more assumed to be degenerate
and particle-hole invariant. As before, the impurities are for convenience only coupled through a spin-exchange
and not by hopping terms, which we will take into account as perturbations.

\subsection{CFT preliminaries for the tetramer}
\label{CFT preliminaries for the tetramer}

Given our choice of the model (\ref{Ham-tetramer}), the charge degrees of
freedom can be still represented by four independent $SU(2)_1$ CFTs, one for each bath.
Concerning the spin degrees of freedom, the way in which the impurities are exchange-coupled 
naturally leads to the following conformal embedding scheme
\bea
&& \left(SU(2)_1^{(1)} \times SU(2)_1^{(3)}\right)\times
\left(SU(2)_1^{(2)} \times SU(2)_1^{(4)}\right) \nonumber \\
&& \rightarrow
\left(SU(2)_2^{(1-3)} \times Z_2^{(1)}\right)\times
\left(SU(2)_2^{(2-4)} \times Z_2^{(2)}\right)\nonumber\\
&& \rightarrow SU(2)_4 \times Z_2^{(1)}\times Z_2^{(2)}
\times \Bigg[c=1 \left(\mbox{CFT}\right)_{p'=6}\Bigg]
\label{embedding plaquette}
\eea
where $c=1$ CFT stands for the 
$Z_2$ orbifold of a free bosonic CFT (the bosonic field $\phi$ and $-\phi$ must be identified) with compactification radius
$R = \sqrt{2p'}$ and $p'=6$~\cite{DiFrancesco,affleck-2001-594}.
The two step process represented by~\ref{embedding plaquette} corresponds to the
coupling of the $SU(2)_1$ spin sectors of the impurities on the diagonals into $SU(2)_2$,
followed by the coupling of these two new sectors into an $SU(2)_4$. The resulting cosets
are the two Ising sectors and the $c=1$ CFT.
The embedding
can again be rigorously proven through the character decomposition, but the proof is very technical 
so we prefer not to give any detail. 
It is also convenient to represent the two Ising sectors as a
single $c=1$ free bosonic CFT, now with compactification radius
$R = \sqrt{4}$, i.e. $p'=2$.

The $Z_2$ orbifold of a $c=1$ CFT with compactification radius $R=\sqrt{2p'}$ includes, besides the
identity, the following primary fields~\citep[pg.~785]{DiFrancesco}:
\begin{itemize}
\item[(i)] $p'-1$ fields, $\phi_{h}$, with dimensions $h=\lambda^2/4p'$, with $\lambda=1,\dots,p'-1$;
\item[(ii)] a doubly-degenerate field, $\phi^{(a)}_{p'/4}$, $a=1,2$, with dimension $p'/4$;
\item[(iii)] the twist operators $\sigma^{(a)}$ and $\tau^{(a)}$, $a=1,2$, with dimensions
1/16 and 9/16, respectively;
\item[(iv)] the dimension-1 operator $\theta$.
\end{itemize}
The fusion rules among the primary fields can be found for
instance in Ref.~\cite[pg.~786]{DiFrancesco}.

The embedding
\be
SU(2)_2 \times SU(2)_2 \rightarrow SU(2)_4 \times \Bigg[c=1 (\mathrm{CFT})_{p'=6}\Bigg],
\label{embed-Georges}
\ee
has already been discussed in Ref.~\cite{Georges&Sengupta} in the context of the 2-impurity 2-channel Kondo model.
We will see in what follows that some of the anomalies found by those authors do also appear in our model.

\subsection{Fixed points in the tetramer phase diagram}

\begin{figure}
\centerline{\includegraphics[width=12cm]{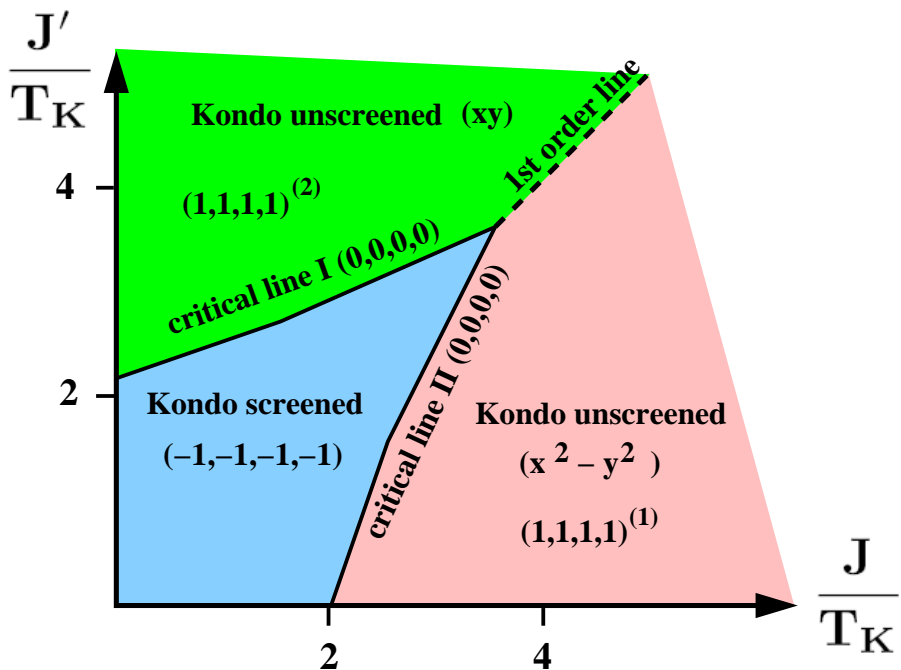}}
\caption{\label{tetramer-phd} Phase diagram of the
impurity tetramer described by the Hamiltonian (\ref{Ham-tetramer}). All phases are discussed in the text, and their 
properties summarized in Table~\ref{Table tetramer}. Since the calculation is numerically heavy, we 
had to use a large $\Lambda=10$, which is sufficient to characterize the low energy spectra, hence the various phases, but 
not adequate to provide accurate estimates of the critical points (for instance, when $J=0$, 
we find $(J'/T_k)_*\simeq 0.24$ instead of the two-impurity value $\simeq 2$~\cite{Jones89}). 
However, even though the absolute values of the critical $J/T_K$ and $J'/T_K$ are underestimated, 
we still believe that the overall shape of the phase diagram should be representative. For this reason, 
we have decided to plot the NRG data rescaled in such a way that, when $J=0$, 
the critical point has the value found in the two-impurity model.~\cite{Jones89}}
\end{figure}

In Fig.~\ref{tetramer-phd}, we draw the phase diagram of
(\ref{Ham-tetramer}) as obtained by NRG. As before, each fixed point
is identified by the $S$-matrices $(S_1,S_2,S_3,S_4)^{(n)}$, where
the superscript $(n)$ is introduced to distinguish between
different fixed points with the same $S$-matrices. 
In Table~\ref{Table tetramer} the main physical properties of the different phases are summarized.

\begin{table}[htb]
\caption{Summary of the main physical properties of the different phases in Fig.~\ref{tetramer-phd}.  
$\rho(0)$ is the zero-frequency DOS of any of the four impurities,  $\rho_0$ being 
its non-interacting value. As in Table~\ref{Table trimer}, several single-particle 
operators are considered, specifying whether they are relevant, in which case their dimension is 
indicated, or not. 
$d^\dagger_i d^\pdag_{i+1}$ and $d^\dagger_i d^\dagger_{i+1}$ denotes hopping and singlet-pairing, 
respectively, along the sides of the plaquette, $i - i+1$ meaning 
$1-2$, $2-3$, $3-4$ or $4-1$.  $d^\dagger_i d^\pdag_{i+2}$ and $d^\dagger_i d^\dagger_{i+2}$  
have the same meaning but along the diagonals, i.e. $i - i+2$ stems for $1-3$ and $2-4$. $x$ is a 
parameter (not to be confused with the coordinate $x$ that is used in the first row to identify the different phases, according to 
Fig.~\ref{tetramer-phd}) that changes continuously along the critical lines I and II, reaching at the first order point 
along the diagonal $J=J'$ the value $x=1$ for line I and $x=2$ for line II.  \\[1.5ex]
\label{Table tetramer}}
\begin{tabular}{@{}llllll}
\br
~~~~ &  $(x^2-y^2)$ & line II & screened & line I & $(xy)$ \\
\mr
$\rho(0)/\rho_0$ & 0 & 1/2 & 1 & 1/2 & 0 \\
$\mathbf{S}_1-\mathbf{S}_2+\mathbf{S}_3-\mathbf{S}_4$ & NOT & $1/3 + x^2/6$ & NOT & NOT & NOT \\
$\mathbf{S}_1 - \mathbf{S}_3$ & NOT & NOT & NOT & $1/2 + x^2/2$ & NOT \\
$\mathbf{S}_2 - \mathbf{S}_4$ & NOT & NOT & NOT & $1/2 + x^2/2$ & NOT \\
$d^\dagger_i d^\pdag_{i+1}$ & NOT & $5/8 + (1-2x)^2/24$ & NOT & NOT & NOT \\
$d^\dagger_i d^\dagger_{i+1}$ & NOT & $5/8 + (1-2x)^2/24$ & NOT & NOT & NOT \\
$d^\dagger_i d^\pdag_{i+2}$ & NOT & NOT & NOT & $1/2 + x^2/2$ & NOT \\
$d^\dagger_i d^\dagger_{i+2}$ & NOT & NOT & NOT & $1/2 + x^2/2$ & NOT \\
\br
\end{tabular}
\end{table}
\noindent \textsl{\bf 1.\quad $\mathbf{(S_1,S_2,S_3,S_4) = (-1,-1,-1,-1)}$}
\\[1.5ex]
This fixed point corresponds to a perfectly Kondo screened phase.
It occurs when $T_K$ is large compared with both $J$ and
$J'$. Once again, we will use the Kondo screened $(-1,-1,-1,-1)$ fixed
point as the ancestor BC to generate all the others through
fusion.
\\[1.5ex]
\noindent \textsl{\bf 2.\quad $\mathbf{(S_1,S_2,S_3,S_4) = (1,1,1,1)^{(1)}}$}
\\[1.5ex]
If $J'=0$ and $J\gg T_K$, the tetramer locks into a non-degenerate
singlet state which is obtained by coupling sites 1 and 3 into a
triplet, as well as sites 2 and 4, and coupling the two triplets
into an overall singlet. This configuration transforms like the function $x^2-y^2$ under the $C_4$ point-group
of the plaquette, hence the label in Fig.~\ref{tetramer-phd}. The singlet decouples from the
conduction electrons which do not feel the presence of the
impurities anymore, resulting in a phase-shift $\delta=0$ in every channel. This phase is Fermi-liquid-like 
and remains stable even in the presence of a finite $J'$, provided the lowest
excitation gap from the ground state of the isolated tetramer is
much larger than $T_K$. Within CFT, there are several possible
fusions which lead the $(-1,-1,-1,-1)$ fixed point to this new
one. One is for instance the fusion with the primary
field $\theta$ of the $c=1$ CFT with $p'=6$.
\\[1.5ex]
\noindent \textsl{\bf 3. \quad $\mathbf{(S_1,S_2,S_3,S_4) = (1,1,1,1)^{(2)}}$}
\\[1.5ex]
If $J=0$, sites 1 and 3 are decoupled from sites 2 and 4, hence the tetramer
reduces to two independent Kondo dimers. If $J'\gg T_K$ each pair of impurities,
1 and 3 or 2 and 4, is strongly bound into a singlet which decouples from the
conduction electrons. This impurity configuration transforms like $xy$ under the $C_4$ point group,
which explains the label in Fig.~\ref{tetramer-phd}.
This fixed point is obviously stable to a small $J$ being turned on, hence,
in analogy with the dimer, it should be obtainable by the $(-1,-1,-1,-1)$ fixed point upon
fusion with $\epsilon_{I}^{(1)}\,\epsilon_{I}^{(2)}$, where $\epsilon_I^{(a)}$, $a=1,2$,
are the energy operator of the two Ising CFTs. This is also the dimension-1
primary field $\theta$ of the $c=1$ CFT with $p'=2$.
The NRG spectrum agrees with this prediction
not only for small $J$, but for the whole region $J'>J$ with $J'\gg T_K$, see
Fig.~\ref{tetramer-phd}.
\\[1.5ex]
\noindent \textsl{\bf 4. \quad First order line}
\\[1.5ex]
If $J=J'\gg T_K$, the tetramer locks into a doubly degenerate spin-singlet, the states
with symmetry $x^2-y^2$ and $xy$ previously mentioned.
The Kondo exchange provides a coupling between these two
configurations only at second order, namely via a quartic conduction-electron
operator, which is therefore irrelevant. Hence the tetramer decouples asymptotically
from the conduction baths, and its degeneracy remains untouched. This is confirmed by
the NRG calculation, which shows the same Fermi liquid spectrum as in the absence of
the impurity-cluster apart from each state being doubly degenerate. This phase is
the analogous of a first order line, hence its name in
Fig.~\ref{Fig-tetramer}, with a relevant operator of dimension 0 that
describes the splitting of the double degeneracy of the tetramer.
\\[1.5ex]
\noindent \textsl{\bf 5. \quad $\mathbf{(S_1,S_2,S_3,S_4) = (0,0,0,0)^{(1)}}$ and $\mathbf{(0,0,0,0)^{(2)}}$}
\\[1.5ex]
The Kondo screened phase at small $J$ and $J'$ is essentially different from the
two unscreened phases at large $J>J'$ and large $J'>J$, respectively.
Hence there are two critical lines that start from the $J'=0$ axis as well as
from the $J=0$ one, see Fig.~\ref{tetramer-phd}, and finally merge with the first order line
at large $J=J'$. This might happen either through a multicritical point or by a gradual evolution
of each line into a first order critical point. The latter scenario is actually realized, since,
unlike in the trimer model, the NRG low-energy spectra along both critical lines
varies continuously, signaling the existence of marginal perturbations.
For the same reason, a precise identification of these critical lines with appropriate boundary CFTs is
not a simple task.

Let us start from the simplest case at $J=0$, which corresponds to two independent dimers, sites 1 plus 3
and sites 2 plus 4. The fixed point which separates the Kondo screened phase from the unscreened one is
obviously the superposition of the fixed points of each dimer, discussed previously.
It is obtained by the Kondo screened fixed point upon
fusion with the product $\sigma_{I}^{(1)}\,\sigma_{I}^{(2)}$ of the two Ising CFTs~\cite{Affleck92PRL},
and is identified by zero scattering matrices, hence $(0,0,0,0)^{(2)}$ in Fig.~\ref{tetramer-phd}, as well as by a 
residual $\ln 2$ entropy.
We already showed that the fixed point of a dimer can be destabilized only by the symmetry-breaking operators
(\ref{dimer-AF})-(\ref{dimer-tperp}), which are not generated by a small $J$.
Therefore, a finite $J\ll J'$ does not spoil the unstable fixed point $(0,0,0,0)^{(2)}$, but only
moves its position to larger $J'$, as it generates a weak ferromagnetic exchange along each dimer.
We notice that, upon double fusion,
\ba
&& \Bigg(\sigma_{I}^{(1)}\,\sigma_{I}^{(2)}\Bigg) \times \Bigg(\sigma_{I}^{(1)}\,\sigma_{I}^{(2)}\Bigg) =
I + \epsilon_{I}^{(1)} + \epsilon_{I}^{(2)} + \epsilon_{I}^{(1)}\,\epsilon_{I}^{(2)} \\
&& \equiv I + \phi_{1/2} + \theta,
\ea
where the last expression on the right-hand side is written in terms of the corresponding fields
of the $p'=2$, $c=1$ CFT. In agreement with NRG, the operator content includes the marginal
operator $\theta$, besides the dimension-1/2 relevant operator that moves away from the fixed point.
Since for two independent dimers we do know that there is no such marginal operator at the unstable fixed point,
we must conclude that $\theta$ acquires a finite coupling-constant only for $J \neq 0$. This situation, which is quite
exceptional in impurity models, resembles that found in Ref.~\cite{Georges&Sengupta} in the 2-impurity 2-channel Kondo model.
An important discovery of Ref.~\cite{Georges&Sengupta} was that this marginal operator not only
influences the spectrum but also the operator content.
Specifically, Georges and Sengupta recognized, by abelian bosonization of the model, that the fixed point Hamiltonian
in the presence of the marginal operator is similar to an X-ray edge problem in bosonization
language~\cite{Schotte}. Therefore any operator which
involves creation or annihilation of the corresponding ``core-hole'' acquires an additional dimension.
It is not difficult to realize that the same happens in our case. Indeed the dimension-1/2 relevant operator
can be mapped within bosonization~\cite{Sasha} into the operator
\be
\epsilon_{I}^{(1)} + \epsilon_{I}^{(2)} \rightarrow d^\dagger\,\Psi(0) + \Psi(0)^\dagger\,d,
\label{X-ray-1}
\ee
which represents the creation (annihilation) of a core-electron, $d^\dagger$($d$), and the contemporary
annihilation (creation) of a conduction electron at the core-hole site, $\Psi(0)$ $\left(\Psi(0)^\dagger\right)$.
Analogously, the marginal operator transforms like
\[
\epsilon_{I}^{(1)}\,\epsilon_{I}^{(2)} \rightarrow \left(1-d^\dagger\,d\right)\;\Psi(0)^\dagger\,\Psi(0),
\]
which corresponds to the interaction between the core-hole and the conduction electrons.
In the presence of this term, the dimension of the relevant operator (\ref{X-ray-1}) changes according to
\[
\frac{1}{2} \rightarrow \frac{(2-2x)^2}{8},
\]
where $x$ parametrizes the critical line, and is actually related to the phase-shift induced by the core-hole
in the equivalent X-ray edge problem. Since the end-point at $J=J'$ is expected to be a first order one,
we conclude that $x$ moves from 0, $J=0$, to 1, $J=J'$, along the critical line. The dimensions of all the other
dimension-1/2 symmetry-breaking operators changes instead as
\[
\frac{1}{2} \rightarrow \frac{1}{2} + \frac{(2x)^2}{8},
\]
so that they all become marginal at $J=J'$. Notice that, since the twist operators are not affected by the
marginal perturbation, the $S$-matrices do not change along the line.

Concerning the other critical line, $(S_1,S_2,S_3,S_4) = (0,0,0,0)^{(1)}$ in Fig.~\ref{tetramer-phd},
we find that, when $J'=0$, the NRG spectrum is reproduced with a good approximation by fusing the 
$(-1,-1,-1,-1)$ BC with the primary field $\phi_{1/6}$ of the
$p'=6$, $c=1$ CFT. Once again, since $\phi_{1/6}\times \phi_{1/6} = I + \theta + \phi_{2/3}$, the operator content
includes, besides the relevant operator of dimension 2/3 that moves away from criticality,
the marginal operator $\theta$ of the $p'=6$, $c=1$ CFT, which explains 
the continuous evolution of the NRG spectrum from $J'=0$ to $J'=J$. The role of this marginal operator should be
similar to its analogous on the other critical line. Therefore, following the previous analysis and
in accordance with Ref.~\cite{Georges&Sengupta}, we expect the critical line to be parametrized by a
``phase-shift'' $x$ that modifies not only the spectrum but also the operator content. In particular, the
dimension of the operator $\phi_{2/3}$ that moves away from the critical line changes according to
\[
\frac{2}{3} \rightarrow \frac{(4-2x)^2}{24}.
\]
Since this operator eventually acquires vanishing dimension at $J'\to J$, we conclude that $x\to 2$ at the
end-point. The precise determination of $x$ along the line is however difficult to extract from our NRG spectra.

At $x=0$, the most relevant symmetry breaking operator
would correspond to the staggered magnetization
\be
\boldsymbol{J}_1-\boldsymbol{J}_2+\boldsymbol{J}_3- \boldsymbol{J}_4,
\label{tetramer-Ms}
\ee
with dimension 1/3, which, at finite $x$, changes into
\[
\frac{1}{3} + \frac{(2x)^2}{24},
\]
hence becomes marginal at $J=J'$. This is physically sound, since at $J=J'$ there is maximum spin frustration.
Besides the staggered magnetization, there are other less relevant
symmetry-breaking operators of dimension 2/3 at $x=0$. One of them is the singlet four-fermion operator
\[
\big(\boldsymbol{J}_1 - \boldsymbol{J}_3\big)\cdot
\big(\boldsymbol{J}_1 - \boldsymbol{J}_2 + \boldsymbol{J}_3
- \boldsymbol{J}_4\big) + (1,3)\leftrightarrow (2,4),
\]
whose dimension at $x\not =0$ is $1/2 + (2+2x)^2/24$, thus becoming soon irrelevant.

The other symmetry-breaking operators of dimension 2/3 correspond actually to all possible
mean-field decoupling schemes of the exchange term
\[
\big(\boldsymbol{J}_1 + \boldsymbol{J}_3\big)\cdot
\big(\boldsymbol{J}_2 +  \boldsymbol{J}_4\big),
\]
into inter-bath single-particle operators. Among them, we just mention the inter-bath hopping,
\[
\sum_\sigma\, \big(c^\dagger_{1\sigma} + c^\dagger_{3\sigma}\big)
\big(c^\pdag_{2\sigma} + c^\pdag_{4\sigma}\big) + H.c.,
\]
as well as the $d$-wave Cooper pairing,
\[
\big(c^\dagger_{1\uparrow}-c^\dagger_{3\uparrow}\big)
\big(c^\dagger_{2\downarrow} - c^\dagger_{4\downarrow}\big) -
\big(\uparrow \, \leftrightarrow \, \downarrow\big).
\]
They are all degenerate and, at $x\not = 0$, have dimension
\[
\frac{5}{8} + \frac{(1-2x)^2}{24}.
\]
At $x=2$ they become marginal, but, interestingly enough, their dimension is non-monotonic in $x$,
although always greater than the staggered magnetization (\ref{tetramer-Ms}). Along this line, too,
the $S$-matrices and the residual entropy remain constant and equal to  $(S_1,S_2,S_3,S_4) = (0,0,0,0)$ and $\ln 2$, respectively.

Like in the dimer and trimer examples, the relevance of the inter-bath hopping implies that 
both critical lines are no more accessible if, instead of four impurities coupled by a spin-exchange,
one considers four impurities coupled by a single particle hopping, which is the actual situation
within cluster-DMFT. Unfortunately, in this case we cannot obtain reliable spectral functions by NRG because of numerical
limitations. Therefore, we cannot verify whether, in spite of the fact that the critical point
is washed out, a sizable critical region still survives. However we tend to believe that it is the case,
just like in the previous examples.  Finally, since both critical lines are stable towards the conventional
particle-hole symmetry breaking, the phase diagram for $J\not = J'$, as function of $U/\Gamma$ and of
the average impurity-occupancy, still looks like the dimer one, see Fig.~\ref{dimerpdmu}.

\section{Discussion and conclusions}

In this Topical Review, we attempted to uncover the key features that distinguish Anderson impurity clusters from
single-impurity models and that could play an important role within cluster dynamical mean-field theory as opposed to
its original single-site formulation.
All the examples that we have studied, namely 2-, 3- and 4-impurity clusters, share very similar properties.

In particular, if the impurities within the cluster are coupled to one another by a two-body spin-exchange 
while each of them is hybridized with
its own separate conduction bath, the phase diagrams as function of the average impurity occupancy and of
the Hubbard $U$ are practically the same, see Fig.~\ref{dimerpdmu}. For $U/\Gamma$ less than a critical value, perfect
Kondo screening occurs and the impurity-cluster spectral functions show the conventional Kondo resonance.
Above that critical value, the inter-impurity exchange prevails instead and takes care of freezing out the impurity degrees
of freedom. Here, the impurity spectral functions develop a pseudo-gap at 
the chemical potential, which is gradually filled in by ``doping'', i.e., by
moving the average impurity occupancy away from half-filling. These two regimes are separated by a critical line
that is identified by several instability channels. In all cases, the instability channels correspond to
all possible mean-field decoupling schemes of the spin-exchange into bilinear operators. They include the
intra-bath magnetization, staggered according to the signs and relative strengths of the spin-exchange constants, 
and all singlet inter-bath bilinear operators, like the inter-bath hoppings or singlet Cooper pairs.
The trend from the dimer towards the tetramer is towards a prevailing instability in the staggered magnetization
channel. The dynamics across the critical point is basically controlled by \underline{two} separate energy scales. One of them, which we
denoted as $T_+$, is finite across the transition and roughly of the order of the spin-exchange.
The other, $T_-$, is the scale generated by the deviation $X$ from the critical line. It vanishes 
as $T_- \sim |X|^\alpha$, where, in the most interesting case of the tetramer, the exponent
$1\leq \alpha \leq 3$, is
non-universal and depends on the frustration. From the point of view of the impurity spectral functions,
see e.g. Fig.~\ref{DOSexE},
$T_+$ is the width of a broad incoherent peak within the Hubbard side-bands, smooth across the critical point.
On the contrary, $T_-$ is the width of the Kondo-like resonance that develops on top of the broader one, in the
Kondo screened phase. As the critical point is approached, the Kondo resonance becomes narrower, and
eventually disappears right at the critical point, leaving behind only the incoherent peak. In the
unscreened phase, $T_-$ controls the width of the pseudo-gap that opens inside the incoherent part.

If the impurities inside a cluster are coupled one another by a single-particle hopping, $t_\perp$, instead of a spin-exchange, the transition
turns into a crossover from the Kondo-resonance behavior to the pseudo-gaped one. Indeed, the
inter-impurity hopping plays a double role. On the one hand it generates, for large $U$, a spin-exchange, $J=4t_\perp^2/U$,
that might drive the model across the critical point. On the other hand, it also induces
a small inter-bath hybridization, $V\sim J_K\, t_\perp/U$, that is a relevant perturbation and destabilizes 
the critical point. 
Specifically, $V$ cuts off all critical point singularities below an energy scale $E_{cut-off}\sim V^\beta$, where
$\beta\geq 3$ in the tetramer and $\beta=2$ in the dimer. Since $V$ is small
and $\beta$ large, we expect that, in spite of the critical point
being no longer accessible, a ``critical region'' should still survive if $E_{cut-off}\ll T_+\sim J$.
We indeed found evidences in favor of this scenario both in the dimer as well as in the trimer, where
the impurity spectral functions are numerically accessible.

\bigskip

Coming back to our original scope, let us imagine that we implement a cluster-DMFT simulation of a Hubbard model
using a dimer, a trimer or a tetramer as representative clusters. As $U$ increases driving the model towards
the Mott transition, the effective impurity cluster must necessarily go through the
above mentioned critical region (even if a true criticality is, rigorously speaking, not accessible since
the impurities as well as the baths are coupled by a single-particle hopping).
In this region, the instability channels of the avoided critical point will be amplified and, after the DMFT self-consistency,
can induce a true bulk instability in the lattice model. At half-filling, our results on the impurity clusters suggest
that most likely magnetism appears, even in the presence of frustration. However, since instabilities in particle-hole channels
are weakened or removed by doping away from commensurate fillings, while that weakening does not happen in particle-particle channels, 
a superconducting dome may emerge near half-filling. Indeed, recent cluster DMFT simulations of the
Hubbard model on a square lattice~\cite{Jarrell-2005,Capone-2006} found evidence of a $d$-wave superconducting phase
away from half-filling, in close analogy with the phase diagram of high-$T_c$ superconductors.

Nevertheless, irrespectively of which symmetry-broken phase actually occurs at low temperatures, the physics of 
impurity-clusters suggests that, in the normal phase above a critical temperature, the transition to the Mott insulator is accompanied
by the gradual opening of a pseudo-gap in the single-particle spectral function. In this pseudo-gaped region, Fermi-liquid
behavior, i.e. $\Sigma(i\omega) \sim i\omega$, is recovered only at extremely low energies, suggesting 
the existence of a finite temperature non-Fermi liquid behavior. Evidence in favor of this scenario has been found in a 
recent cluster-DMFT simulation of a paramagnetic two-dimensional Hubbard model.~\cite{imada-2007} 

Another aspect worth emphasizing concerns the behavior of the Drude weight in the metal away from half-filling 
across the symmetry-breaking phase-transition. From the point of view of the effective impurity model, a 
symmetry breaking in the conduction baths opens up new screening channels that can rid the impurity of its residual entropy
at the critical point. This in turns leads to an increase of screening energy gain that,
translated back into the lattice model, implies 
an \underline{increase} of band-energy gain, i.e. of the Drude weight. This behavior is actually the fingerprint of this
kind of instability that reflects the underlying impurity critical point,
as opposed to the conventional Stoner- or BCS-instability that are accompanied by a \underline{decrease} of Drude weight.
The increase of Drude weight has been indeed observed in DMFT simulations of the two-dimensional
Hubbard model~\cite{Jarrell-2005} as well as of a two-band Hubbard model~\cite{Capone04} that maps, within DMFT,
onto the impurity dimer.

\ack
We acknowledge helpful discussions with E. Tosatti. We are very grateful to A. Georges, who pointed to our attention
Ref.~\cite{Georges&Sengupta} that has been enlightening for our study.

\appendix

\section*{Appendix: CFT at work}
\addcontentsline{toc}{section}{Appendix: CFT at work}
\setcounter{section}{1}

In this Appendix, we show how {\sl conformal embedding} works in a few simple examples.
We do it through the identification of the partition function, 
the so-called {\sl character decomposition}.

The first step of bosonization in one dimension is the linearization of the free-electron spectrum around the 
Fermi momentum~\cite{Sasha}.
This linearization is not expected to affect the low energy behavior provided the perturbations are weak compared 
to the band-width. Therefore let us consider, instead of a tight-binding model, free spinless Dirac fermions, 
which have indeed a linear spectrum, on a chain of length $L$ with anti-periodic boundary conditions. 
We are going to consider Dirac fermions moving only in one direction, namely with 
a single chirality, because this is the case relevant to a semi-infinite chain where  
the single-particle wave-functions with negative 
momenta are not independent from those with positive ones. 

The single-particle wave-functions for a chiral Dirac fermion are plane waves with momentum 
\[
k = \frac{\pi}{L}\,(2n-1),
\]
$n$ being integer. The Hamiltonian in momentum space reads
\be
\mathcal{H} = v_F\,\sum_k\, k\, c^\dagger_k\,c^\pdag_k,
\ee
where $v_F$ has to be identified with the Fermi velocity of the original tight-binding model. 
Let us define, for positive $k$,
\ba
\alpha^\pdag_k &=& c^\pdag_k,\\
\beta^\pdag_k &=& c^\dagger_{-k},
\ea
so that, apart from an (actually infinite) constant, the Hamiltonian becomes
\be
\mathcal{H} = v_F\,\sum_{k>0}\, k\, \left(\alpha^\dagger_k\,\alpha^\pdag_k + \beta^\dagger_k\,\beta^\pdag_k\right). 
\ee
The partition function at temperature $T$ is simply
\be
Z_{Dirac} = \prod_{k>0}\,\bigg[1+\exp\left(-\beta\,v_F\,k\right)\bigg]^2 = 
\prod_{n\geq 1}\, \left(1 + q^{n-1/2}\right)^2,
\ee
where conventionally~\cite{DiFrancesco} $q$ is defined as 
\[
q = \exp\left(-\beta\,\frac{2\pi v_F}{L}\right) \equiv {\rm e}^{2\pi i \tau}.
\]  
One can show that 
\be
Z_{Dirac}(q) = \frac{\theta_3(q)}{\varphi(q)},
\label{NRG-CFT:ZDirac}
\ee
where $\theta_3$ is the third Jacobi theta-function 
\[
\theta_3(q) = \sum_{n=-\infty}^\infty\, q^{n^2/2},
\]
and $\varphi$ the Euler function
\[
\varphi(q) = \prod_{n\geq 1}\,\left(1-q^n\right).
\]

\bigskip

On the other hand, it is known by bosonization~\cite{Sasha} that, for positive $p=2\pi n/L>0$, the operators
\[
b^\pdag_p = -i\sqrt{\frac{2\pi}{pL}}\,\rho(p) =-i\sqrt{\frac{2\pi}{pL}}\,\sum_k \,c^\dagger_k\,c^\pdag_{k+p}, \qquad 
b^\dagger_p = i\sqrt{\frac{2\pi}{pL}}\,\rho(-p),
\]
satisfy bosonic commutation relations. In addition their equation of motion can be reproduced by the 
Hamiltonian 
\be
\mathcal{H} = \frac{\pi v_F}{L} \,\sum_p\, \rho(p)\,\rho(-p) = 
v_F\,\sum_{p>0}\, p\,b^\dagger_p\,b^\pdag_p + \frac{\pi v_F}{L}\,\Delta\,N^2,
\ee
where it is assumed that $\rho(p=0)$ is the variation $\Delta\, N$ of the electron number with respect to a reference value. 
The partition function of this bosonic model is the product of the free-boson term 
\[
\prod_{p>0}\,\left[1 - \exp\left(-\beta\,v_F\,p\right)\right]^{-1} = \prod_{n>0}\,\left(1-q^n\right)^{-1} 
= \varphi(q)^{-1}
\]
plus the contribution of $\Delta\,N$ which is, assuming an infinite reference number,  
\[
\sum_{n=-\infty}^\infty\, \exp\left(-\beta\,\frac{v_F\pi}{L}\,n^2\right) = 
\sum_{n=-\infty}^\infty\, q^{n^2/2} = \theta_3(q).
\]
One immediately recognizes that the bosonic partition function coincides with the fermionic one.

\bigskip

We can proceed further on, and consider spinful Dirac fermions. Obviously, the partition function is the square of 
$Z_{Dirac}$ in Eq.~(\ref{NRG-CFT:ZDirac}). As the simplest example of {\sl conformal embedding} and 
{\sl character decomposition}, we consider a perturbation that only preserves 
independently the spin $SU(2)$ symmetry and the charge isospin $SU(2)$, defined through the generators $\mathbf{I}$, 
which are the $q=0$ components of the so-called isospin current operators   
\ba
I_z(q) &=& \frac{1}{2}\,\sum_{k \sigma}\, \Big(c^\dagger_{k\sigma}c^\pdag_{k+q\sigma} - \delta_{q 0 }\Big),\\
I^+(q) &=& \sum_{k}\, c^\dagger_{k\uparrow}\,c^\dagger_{-k-q\downarrow},\\
I^-(q) &=& \Big(I^+\Big)^\dagger.
\ea
The spin $SU(2)$ current operators are instead 
\be
\mathbf{S}(q) = \frac{1}{2}\sum_{k \alpha\beta}\, c^\dagger_{k\alpha}\,\boldsymbol{\sigma}_{\alpha\beta}\, 
c^\pdag_{k+q\beta},
\ee
where $\boldsymbol{\sigma}$ are the Pauli matrices. In real space these current operators, $J_a=I_a,\,S_a$, 
satisfy the commutation relations
\[
\bigg[J_a(x),J_b(y)\bigg] 
= i\epsilon_{abc}\,\delta(x-y)\,J_c(x) - i\,k\,\frac{1}{4\pi}\,\delta_{ab}\, \frac{\partial \delta(x-y)}{\partial x},
\]
with $k=1$. For generic $k\geq 1$, the above commutation relations identify 
an $SU(2)_k$ CFT~\cite{DiFrancesco},
where the label $k$ may be regarded as the number of channels which are used to build up the generators. 
An $SU(2)_k$ CFT has primary fields $\boldsymbol{\phi}^{(k)}_{2j}$ with spin quantum numbers 
$j$, such that $2j=0,1,\dots,k$. Their dimensions are $x_j = j(j+1)/(k+2)$. 
The product of two primary fields with spin $j$ and $j'$ yields all primary 
fields with spin between $|j-j'|$ and $\mathrm{min}(k-j-j',j+j')$. The Hilbert space of the theory is obtained by applying the 
primary fields on the reference vacuum state and, from this ancestor state, by generating all descendant 
states applying the current operators 
with $q<0$. This is what is called a {\sl conformal tower}~\cite{DiFrancesco}.
The energy difference between the descendant states and 
their ancestor is an integer multiple of the fundamental level spacing $\Delta = 2\pi v_F/L$. 
The character $\chi^{(k)}_{2j}$ represents the 
contribution to the partition function of the conformal tower generated by the primary field 
$\boldsymbol{\phi}^{(k)}_{2j}$~\cite{DiFrancesco}.

This construction may look abstruse but actually has a simple physical interpretation. Let us consider again a single 
spinful fermion, $k=1$. 
Let us further assume that on average the number of electrons is equal to the number of sites, and the latter is even. 
In this case the ground state is obtained by filling with two 
electrons of opposite spin all single-particle levels below the chemical potential, which lies in the middle of 
two consecutive single-particle levels separated by the fundamental spacing $\Delta$, from now on our energy unit.   
With this definition, the Hilbert space can be constructed as follows. One can start from the vacuum and act on 
it with particle-hole excitations, namely with $I_z(q)$ or $\mathbf{S}(q)$ with $q<0$. In addition, one can apply 
the operators $I^+(q)$ or $I^-(q)$, again with $q<0$, 
to change by an even multiple the number of electrons, and then consider all particle-hole excitations 
on top of these states. In this way, one obtains all states which have even number of electrons, like the vacuum state. 
This is nothing but the conformal towers  
obtained by the ancestor fields $\boldsymbol{\phi}^{(1)}_{0}$  
in both the charge and spin $SU(2)_1$ sectors, which should therefore contribute to the 
partition function with the product of characters $\left(\chi^{(1)}_0\right)_{charge}\,\left(\chi^{(1)}_0\right)_{spin}$. 

The rest of the Hilbert space includes all states with odd number of electrons. Since a single electron carries 
isospin and spin 1/2, all these states have half-odd integer values of $I_z$ and $S_z$. 
One realizes that they can all be obtained by applying the 
product of the isospin and spin primary fields $\phi^{(1)}_{1\, charge}\times \phi^{(1)}_{1\, spin}$, 
which is nothing but the single electron operator, and construct 
out of it all descendant states. Their contribution to the partition function should then be 
$\left(\chi^{(1)}_1\right)_{charge}\,\left(\chi^{(1)}_1\right)_{spin}$. The expression of the $SU(2)_k$ 
characters~\cite[page 586]{DiFrancesco} is 
\[
\chi^{(k)}_l(q) = \frac{1}{\eta(q)^3}\, \sum_{n=-\infty}^\infty\, \bigg[2n\left(k+2\right)+l+1\bigg]\, 
q^{(2n(k+2)+l+1)^2/4(k+2)},
\]
where 
\[
\eta(q) = q^{1/24}\,\varphi(q),
\]
is the Dedekind function. In the specific case of $k=1$,  
\ba
\chi^{(1)}_0(q) &=& \sqrt{\frac{\theta_3(q)^2 + \theta_4(q)^2}{2\eta(q)^2}},\\
\chi^{(1)}_1(q) &=& \sqrt{\frac{\theta_3(q)^2 - \theta_4(q)^2}{2\eta(q)^2}},\\
\ea
where 
\[
\theta_4(q) = \sum_{n=-\infty}^\infty \, (-1)^n\, q^{n^2/2},
\]
is the fourth Jacobi theta-function. Hence we find that 
\bea
\left(\chi^{(1)}_0\right)_{charge}\,\left(\chi^{(1)}_0\right)_{spin} + 
\left(\chi^{(1)}_1\right)_{charge}\,\left(\chi^{(1)}_1\right)_{spin}
&=& \frac{\theta_3(q)^2}{\eta(q)^2} \nonumber \\
&=& q^{-1/12}\, Z_{Dirac}(q)^2,
\label{CFT-Z-single-channel}
\eea
which, apart from the vacuum polarization contribution $q^{-1/12}$, is exactly the partition function 
of two species of Dirac fermions. 
The spectrum, and correspondingly the partition function, can be represented as in Table~\ref{single-channel}.
\begin{table}[htb]
\caption{The spectrum of spinful electrons when the ground state contains an even number of particles.
\label{single-channel}}
\begin{indented}
\item[] \begin{math}
\begin{array}{lll}\br
I & S & x \\ \mr
0 & 0 & 0 \\ \mr
1/2 & 1/2 & 1/2 \\ \br
\end{array}
\end{math}
\end{indented}
\end{table}
In that and all forthcoming tables, we identify each conformal tower by the quantum numbers of the primary fields 
which generate the ancestor states. $x$ is the energy in units of $\Delta$ of the ancestor state with respect to the 
chemical potential. The descendant levels of an ancestor 
have energies $x + n$, with $n$ a positive integer. Notice that an important consequence of conformal invariance is that 
the energy of each state in units of $\Delta$ coincide with the dimension of the operator which, 
applied to the vacuum, yields that state.    

Following the same reasoning, one can easily show that the table of the energy spectrum in the case of 
odd chains at half-filling (i.e. odd average number of electrons)  
is the one of Table~\ref{single-channel-odd}. 
\begin{table}[htb]
\caption{The spectrum of spinful electrons when the ground state contains an odd number of particles. 
\label{single-channel-odd}}
\begin{indented}
\item[] \begin{math}
\begin{array}{lll}\br
I & S & x - 1/4 \\ \mr
1/2 & 0 & 0 \\ \mr
0 & 1/2 & 0  \\ \br
\end{array}
\end{math}
\end{indented}
\end{table}
The ground state is fourfold degenerate, the chemical potential coinciding with a single-particle 
level. One can readily realize that the even and odd chain spectra can be turned one into the other 
by {\sl fusion} with a charge or a spin primary field $\boldsymbol{\phi}^{(1)}_{1}$.  
This is the physics of the single-channel spin-1/2 Kondo impurity model. Indeed, when Kondo effect is established, 
the impurity site becomes effectively a new site of the chain, thus changing the parity of the number of sites, i.e. 
the boundary conditions.    

\paragraph{The two channel model}

Let us consider a more involved {\sl conformal embedding} in the case of two channels of spinful fermions. 
The partition function is the square of the partition function (\ref{CFT-Z-single-channel}) of a single channel, 
and can be written for even chains as 
\be
Z= \sum_{n_1,n_2=0,1}\, \Bigg(\chi^{(1)}_{n_1}\,\chi^{(1)}_{n_2}\Bigg)_{charge}\; 
\Bigg(\chi^{(1)}_{n_1}\,\chi^{(1)}_{n_2}\Bigg)_{spin},
\ee
where $n_1$ refers to channel 1 and $n_2$ to channel 2. For odd chains, one readily finds that 
\be
Z= \sum_{n_1,n_2=0,1}\, \Bigg(\chi^{(1)}_{n_1}\,\chi^{(1)}_{n_2}\Bigg)_{charge}\; 
\Bigg(\chi^{(1)}_{1-n_1}\,\chi^{(1)}_{1-n_2}\Bigg)_{spin}.
\ee
These expressions manifestly show that the two-channel free conduction electrons are invariant under independent 
spin or isospin $SU(2)$ transformations within each channel, namely under a large 
symmetry $SU(2)\times SU(2) \times SU(2) \times SU(2)$. 

Let us suppose that the impurity couples only to the spin-current operators in such a way that only the overall 
SU(2) spin symmetry is preserved. Therefore, while the charge degrees of freedom can still be represented 
by two $SU(2)_1$ CFT's, the appropriate conformal embedding for the spin sectors should involve an $SU(2)_2$ CFT, 
since the total spin current is made up of two channels, times the coset CFT, namely 
\[
SU(2)_1 \times SU(2)_1 \rightarrow SU(2)_2\,\times\, \frac{SU(2)_1 \times SU(2)_1}{SU(2)_2}.
\]
Since the central charge is conserved and each $SU(2)_k$ has a central charge $3k/(k+2)$, the coset theory should 
have $c=1/2$, which corresponds to the central charge of an Ising CFT. This can be proved rigorously by the 
{\sl character decomposition}. 

The Ising CFT has three primary fields, the identity $I$, with dimension 0, the energy field $\epsilon$, with dimension 1/2, 
and the spin field $\sigma$ with dimension 1/16. The fusion rules are~\cite{DiFrancesco}
\be
I\times I = I,\;\; \epsilon\times \epsilon =I,\;\;
\sigma\times\sigma = I + \epsilon,\;\; I\times\epsilon =\epsilon,\;\;
I\times \sigma = \sigma,\;\; \epsilon\times \sigma = \sigma.
\label{Ising-fusion}
\ee
The characters $\chi^I_x$, where $x$ is the dimension of the primary field, are given by 
(all functions are assumed to depend on the variable $q$, even when not indicated) 
\ba
\chi^I_0 &=& \frac{1}{2}\, \Bigg[ \sqrt{\frac{\theta_3}{\eta}} + \sqrt{\frac{\theta_4}{\eta}}\Bigg],\\
\chi^I_{1/2} &=& \frac{1}{2}\, \Bigg[ \sqrt{\frac{\theta_3}{\eta}} - \sqrt{\frac{\theta_4}{\eta}}\Bigg],\\
\chi^I_{1/16} &=& \sqrt{\frac{1}{2}}\, \sqrt{\frac{\theta_2}{\eta}},\,
\ea
where the second Jacobi function is defined by
\[
\theta_2 = \sum_{n=-\infty}^\infty\, q^{(n-1/2)^2/2}.
\]

One can show that the product of characters of two 
$SU(2)_1$ CFTs, $\chi^{(1)}_{2j}\,\chi^{(1)}_{2j'}$, can be related to the product of characters 
of an $SU(2)_2$ and an Ising CFTs, $\chi^{(2)}_{2l}\,\chi^I_x$, by
\ba
\chi^{(1)}_{0}\,\chi^{(1)}_{0} &=& \chi^{(2)}_{0}\,\chi^I_{0} + \chi^{(2)}_{2}\,\chi^I_{1/2},\\
\chi^{(1)}_{0}\,\chi^{(1)}_{1} &=& \chi^{(1)}_{1}\,\chi^{(1)}_{0} = \chi^{(2)}_{1}\,\chi^I_{1/16},\\
\chi^{(1)}_{1}\,\chi^{(1)}_{1} &=& \chi^{(2)}_{0}\,\chi^I_{1/2} + \chi^{(2)}_{2}\,\chi^I_{0}.
\ea
By means of the above decomposition one can write the tables of the spectra for even and odd chains, as given in 
Table~\ref{k=2-even-odd}.
\begin{table}[thb]
\caption{The spectra of two channels on even chains, left table, and odd chains, right table. 
$I_1$ and $I_2$ are the isospin value of each channel, $S$ the value of the total spin 
and $\mathrm{Ising}$ refers to the Ising primary fields.
\label{k=2-even-odd}}
\begin{indented}
\item[]\begin{math}
\begin{array}{lllll}\br
I_1 & I_2 & S & \mathrm{Ising} & x \\ \mr
0 & 0 & 0 & 0 & 0 \\ \mr
0 & 0 & 1 & 1/2 & 1 \\ \mr
1/2 & 0 & 1/2 & 1/16 & 1/2 \\ \mr
0 & 1/2 & 1/2 & 1/16 & 1/2 \\ \mr
1/2 & 1/2 & 0 & 1/2 & 1 \\ \mr
1/2 & 1/2 & 1 & 0 & 1 \\ \br
\end{array}
\qquad 
\begin{array}{lllll}\br
I_1 & I_2 & S & \mathrm{Ising} & x-1/2 \\ \mr
0 & 0 & 0 & 1/2 & 0 \\ \mr
0 & 0 & 1 & 0 & 0 \\ \mr
1/2 & 0 & 1/2 & 1/16 & 0 \\ \mr
0 & 1/2 & 1/2 & 1/16 & 0 \\ \mr
1/2 & 1/2 & 0 & 0 & 0 \\ \mr
1/2 & 1/2 & 1 & 1/2 & 1 \\ \br
\end{array}
\end{math}
\end{indented}
\end{table}
By these tables one easily realizes that the spectrum of even chains can be turned into that of odd chains, and vice versa,
either by fusion with the $SU(2)_2$ primary field of spin 1, or by fusion with the Ising field $\epsilon$.  
As we mention in Section~\ref{The impurity dimer}, the unstable fixed point of the two spin-1/2 Kondo impurity model 
was found by Affleck and Ludwig~\cite{Affleck92PRL,Affleck95} to correspond to fusion with the Ising field $\sigma$. If one  
performs such a fusion in the spectra of Table~\ref{k=2-even-odd}, one obtains, by means of the fusion rules 
(\ref{Ising-fusion}), a new spectrum 
\begin{table}[hbt]
\caption{Left table: The spectrum of the unstable fixed point of the two spin-1/2 Kondo impurity model.
Right table: The boundary-operator dimensions at the unstable fixed point. The single and double asterisks identify, 
respectively, the symmetry-invariant and symmetry-breaking relevant physical operators.
\label{k=2-fixed-content}}
\begin{indented}
\item[] \begin{math}
\begin{array}{lllll}\br
I_1 & I_2 & S & \mathrm{Ising} & x-1/16 \\ \mr
0 & 0 & 0 & 1/16 & 0 \\ \mr
1/2 & 0 & 1/2 & 0 & 3/8 \\ \mr
0 & 1/2 & 1/2 & 0 & 3/8 \\ \mr
0 & 0 & 1 & 1/16 & 1/2 \\ \mr
1/2 & 1/2 & 0 & 1/16 & 1/2 \\ \mr
1/2 & 0 & 1/2 & 1/2 & 7/8 \\ \mr
0 & 1/2 & 1/2 & 1/2 & 7/8 \\ \mr
1/2 & 1/2 & 1 & 1/16 & 1 \\ \br
\end{array}
\end{math}
\qquad 
\begin{math}
\begin{array}{lllll}\br
I_1 & I_2 & S & \mathrm{Ising} & x \\ \mr
0 & 0 & 0 &       0 & 0 \\ \mr
0 & 0 & 0 &       1/2 & 1/2^{(*)} \\ \mr
1/2 & 0 & 1/2 &   1/16 & 1/2 \\ \mr
0 & 1/2 & 1/2 &   1/16 & 1/2 \\ \mr
0 & 0 & 1 &       0 & 1/2^{(**)} \\ \mr
1/2 & 1/2 & 0 &   0 & 1/2^{(**)} \\ \mr
0 & 0 & 1 &       1/2 & 1 \\ \mr
1/2 & 1/2 & 0 &   1/2 & 1 \\ \mr
1/2 & 1/2 & 1 &   0 & 1 \\ \mr
1/2 & 1/2 & 1 &   1/2 & 3/2 \\ \br
\end{array}
\end{math}
\end{indented}
\end{table}
that was shown to reproduce the NRG spectrum obtained by Jones and Varma~\cite{Jones89}.
By performing a further fusion (so-called {\sl double fusion}~\cite{Cardy,affleck-1995-26}) 
with $\sigma$, one determines the dimensions of the 
boundary operators at the unstable fixed point, shown in the same Table~\ref{k=2-fixed-content}. Their physical meaning 
is discussed in Section~\ref{The impurity dimer}. 


\end{document}